\documentclass[
  twocolumn,
  prb,
  amsmath,
  amssymb,
  superscriptaddress,
]{revtex4}

\usepackage{graphicx}
\usepackage{upgreek}

\newcommand{\s}{\sum\limits}

\newcommand{\pa}{\partial}

\newcommand{\be}{\begin{equation}}
\newcommand{\e}{\end{equation}}
\newcommand{\beml}{\begin{subequations}}
\newcommand{\eml}{\end{subequations}}
\newcommand{\beq}{\begin{eqnarray}}
\newcommand{\eq}{\end{eqnarray}}
\newcommand{\ba}{\begin{array}}
\newcommand{\ea}{\end{array}}
\newcommand{\bpm}{\begin{pmatrix}}
\newcommand{\epm}{\end{pmatrix}}
\newcommand{\bc}{\begin{cases}}
\newcommand{\ec}{\end{cases}}
\newcommand{\lt}{\left}
\newcommand{\rt}{\right}

\newcommand{\la}{\langle}
\newcommand{\ra}{\rangle}
\newcommand{\ep}{\varepsilon}

\newcommand{\bb}{\boldsymbol}

\DeclareMathOperator{\dv}{div}

\begin{document}

\title{Magnetoresistance of compensated semimetals in confined geometries}

\author{P.\,S.~Alekseev}
\affiliation{A.\,F.\,Ioffe Physico-Technical Institute, 194021 St.\,Petersburg, Russia}
\author{A.\,P.~Dmitriev}
\affiliation{A.\,F.\,Ioffe Physico-Technical Institute, 194021 St.\,Petersburg, Russia}
\author{I.\,V.~Gornyi}
\affiliation{Institut f{\"u}r Nanotechnologie, 
Karlsruhe Institute of Technology, 76021 Karlsruhe, Germany}
\affiliation{A.\,F.\,Ioffe Physico-Technical Institute, 194021 St.\,Petersburg, Russia}
\author{V.\,Yu.~Kachorovskii}
\affiliation{A.\,F.\,Ioffe Physico-Technical Institute, 194021 St.\,Petersburg, Russia}
\author{B.\,N.~Narozhny}
\affiliation{\mbox{Institut f\"ur Theorie der Kondensierten Materie,
Karlsruhe Institute of Technology, 76128 Karlsruhe, Germany}}
\affiliation{National Research Nuclear University MEPhI
(Moscow Engineering Physics Institute), 115409 Moscow, Russia}
\author{M.\,Sch{\"u}tt}
\affiliation{School of Physics and Astronomy, 
University of Minnesota, Minneapolis, MN 55455, USA}
\author{M.\,Titov}
\affiliation{Radboud University Nijmegen, 
Institute for Molecules and Materials, NL-6525 AJ Nijmegen, The Netherlands}

\begin{abstract}
Two-component conductors -- e.g., semi-metals and narrow band
semiconductors -- often exhibit unusually strong magnetoresistance in
a wide temperature range. Suppression of the Hall voltage near charge
neutrality in such systems gives rise to a strong quasiparticle drift
in the direction perpendicular to the electric current and magnetic
field. This drift is responsible for a strong geometrical increase
of resistance even in weak magnetic fields. Combining the Boltzmann
kinetic equation with sample electrostatics, we develop a microscopic
theory of magnetotransport in two and three spatial dimensions. The
compensated Hall effect in confined geometry is always accompanied by
electron-hole recombination near the sample edges and at large-scale
inhomogeneities. As the result, classical edge currents may dominate
the resistance in the vicinity of charge compensation. The effect
leads to linear magnetoresistance in two dimensions in a broad range
of parameters. In three dimensions, the magnetoresistance is normally
quadratic in the field, with the linear regime restricted to
rectangular samples with magnetic field directed perpendicular to the
sample surface. Finally, we discuss the effects of heat flow and
temperature inhomogeneities on the magnetoresistance.
\end{abstract}


\maketitle

The theory of magnetotransport in solids \cite{Pippard,dau8} is a mature
branch of condensed matter physics. Measurements of magnetoresistance
and classical Hall effect are long recognized as valuable experimental
tools to characterize conducting samples. Interpreting the experiments
within the standard Drude theory \cite{Pippard,abrikos,Kittel1963},
one may extract many useful sample characteristics such as the
electron mobility and charge density at the Fermi level. However, in
materials with more than one type of charge carriers -- e.g.,
semi-metals and narrow band semiconductors -- the situation is more
complex. Indeed, already in 1928 Kapitsa observed unconventional
magnetoresistance in semi-metal bismuth films \cite{Kapitza1928}. More
recently, interest in magnetotransport has been revived with the
discovery of novel two-component systems including graphene
\cite{graphene1,graphene2,graphene3,graphene4,graphene5,graphene6},
topological insulators \cite{topins1,topins2,topins3,topins4,topins5},
and Weyl semimetals
\cite{weyl1,weyl2,weyl3,weyl4,weyl5,weyl6,weyl7,weyl8,weyl9,weyl10,weyl11}.
A common feature of all such systems is the existence of the charge
neutrality (or, charge compensation) point, where the concentrations
of the positively and negatively charged quasiparticles (electron-like
and hole-like, respectively) are equal and the system is electrically
neutral.

A fast growing number of experiments on novel two-component materials
exhibit unconventional transport properties in magnetic field: (i)
linear magnetoresistance (LMR) was reported in graphene and
topological insulators close to charge neutrality
\cite{Friedman2010,Singh2012,Veldhorst2013,Wang2013,Gusev2013,Weber,Wiedmann2015,Wang2015,Vasileva16}
as well as in narrow-gap semiconductors \cite{Hu2008}, bismuth films
\cite{Yang1999,Yang2000}, and three-dimensional (3D) silver
chalcogenides \cite{Xu1997,Husmann2002,Sun2003} (ii) giant (and
sometimes also linear) magnetoresistance was identified in semimetals
{\it WTe} \cite{WTe1,WTe2,WTe3}, {\it NbP} \cite{NbP1}, {\it LaBi}
\cite{LaBi1,LaSb&LaBi}, {\it{ZrSiS}} \cite{ZrSiS1,ZrSiS2}, multilayer
graphene \cite{mghBN} and many others
\cite{NbSb,ScPtBi,CdAs,NbAs&TaAs,BiTe,LuPdBi}; (iii) finally, the
widely discussed negative magnetoresistance was found in Weyl
semimetals and related materials
\cite{negativeGR,negative1,negative2,negative3,negative4,negative5,negative6,negative7,negative9,negative10,negative11}. Moreover,
negative magnetoresistance may by regarded as a ``smoking gun'' for
detecting a Weyl semimetal \cite{Burkov2015,Sachdev16}, although experiment
\cite{Ganichev2001,Alekseev2016} shows the existence of the effect in
``non-Dirac'' materials as well.

Conventional Drude-like theories of transport in two-component systems
predict parabolic magnetoresistance that saturates in classically
strong fields \cite{Pippard,abrikos,Weiss1954,Gant}. Taking into
account additional relaxation processes may lead to semiclassical
mechanisms of LMR in diverse physical systems including 3D metallic
slabs with complex Fermi surfaces and smooth boundaries
\cite{Azbel,Kaganov}; strongly inhomogeneous or granular materials
\cite{Dykhne,Parish2003,Knap2014,Weber16}; short samples
\cite{Weber16,Yoshida}; disordered 3D metals \cite{Polyakov,Song}; and
compensated two-component systems \cite{Alekseev15}. Quantum effects
result in LMR in strong fields in 3D zero-gap band systems with linear
dispersion \cite{Abrikosov1969,Abrikosov1998,Abrikosov2000}. In weak
fields, resistivity of two-dimensional (2D) electron systems acquires
an interaction correction \cite{ZNA} that is linear in the field.

The extreme quantum limit of
Refs.~\onlinecite{Abrikosov1969,Abrikosov1998,Abrikosov2000} has been
realized in graphene \cite{Friedman2010}, {\it Bi$_2$Te$_3$}
nanosheets \cite{BiTe}, and possibly in the novel topological material
{\it LuPdBi} \cite{LuPdBi}. However, this mechanism is applicable to
the specific case of 3D systems with linear dispersion subjected to a
strong magnetic field ${\hbar\omega_c\gg{T}}$ (as usual, $T$ is the
temperature, $\hbar$ is the Planck constant, and $\omega_c$ is the
electron cyclotron frequency), where all electrons are confined to the
first Landau level. Recently, this approach has been extended to Weyl
semimetals at finite temperatures and with short-range disorder
\cite{Klier}. However, the above conditions are typically not
satisfied by the majority of systems exhibiting unsaturated LMR at
high temperatures.

Experiments on strongly inhomogeneous (or strongly disordered) systems
are often interpreted using the classical approach of
Refs.~\onlinecite{Dykhne,Parish2003}. In particular, the
random-resistor model of Ref.~\onlinecite{Parish2003} was introduced
to explain the non-saturating LMR in granular
materials such as {\it AgSe} \cite{Xu1997,Husmann2002}. More recently,
this mechanism was used to interpret the behavior of the
hydrogen-intercalated epitaxial bilayer graphene
\cite{Weber}. However, this model (as well as the quantum theory of
Refs.~\onlinecite{Abrikosov1969,Abrikosov1998,Abrikosov2000}) does not
distinguish between single- and multi-component systems, contradicting
the crucial role of the charge neutrality point in many aforementioned
experiments. Moreover, both theoretical approaches rely on the
presence of disorder and thus cannot be used to interpret the data
obtained in ultra-clean, homogeneous samples.

A phenomenological theory of magnetotransport in 2D clean,
two-component systems close to charge neutrality was proposed by the
present authors in Ref.~\onlinecite{Alekseev15}. Subjected to a
perpendicular magnetic field, such systems exhibit the compensated
Hall effect, where the Hall voltages due to positively and negatively
charged carriers partially (precisely at charge neutrality --
completely) cancel each other. Such compensation of the Hall voltage
is accompanied by a neutral quasiparticle flow in the lateral
direction relative to the electric current \cite{Titov2013}. In
constrained geometries this leads to a nonuniform distribution of
charge carriers over the sample area, effectively splitting the sample
into the bulk and edge regions. The resistance of the edge region is
dominated by the electron-hole recombination, while the bulk of the
sample exhibits the usual, essentially Drude resistance. The total
resistance of the sample is then obtained by treating the edge and
bulk regions as independent, parallel resistors. The linear dependence
of the sample resistance on the magnetic field arises due to
qualitatively different behavior of the edge region. At charge
neutrality, the resulting LMR persists into the range of classically
strong fields. Away from the neutrality point, the nonzero Hall
voltage leads to the observed saturation of the magnetoresistance.
Similar ideas were recently exploited by some of us to explain the
phenomenon of the giant magnetodrag in graphene
\cite{Titov2013,Narozhny}. The importance of the electron-hole
recombination processes for magnetotransport in narrow-band
semiconductors and semimetals has been pointed out earlier by Rashba
{\em et. al.} in Ref.~\onlinecite{Rashba76}.

In this paper we present a microscopic theory of magnetotransport in
two-component systems. Combining the kinetic equation with
the sample electrostatics, we provide a rigorous justification for the
phenomenological approach of
Ref.~\onlinecite{Alekseev15}. Furthermore, we extend our theory to 3D
systems. We find that although in 3D the magnetoresistance is
typically quadratic in the field, there exists a linear regime in
rectangular samples with magnetic field directed perpendicular to the
sample surface.

The remainder of the paper is organized as follows. First, we discuss
the qualitative physics of magnetotransport in two-component systems.
In the technical part of the paper we present a Boltzmann equation
approach to magnetotransport in finite-size 2D and 3D systems. In the
latter case, we focus on the rectangular sample geometry to simplify
the analysis of the sample electrostatics. We conclude the paper by
discussing the experimental relevance of our results.

\section{Qualitative discussion}
\label{sec:mech}

Let us first recall the results of the classical linear response
theory \cite{Pippard,abrikos,Kittel1963,Weiss1954,Gant} applied to
two-component systems. A system of charge carriers subjected to a
homogeneous external electric field, $\bb{E}$ exhibits an electrical
current. The current density, $\bb{J}=e\bb{j}$, is proportional to the
applied field, ${J_\alpha=\sigma^{\alpha\beta}E_\beta}$, where
$\hat{\sigma}$ is the conductivity tensor. In two-component systems,
one can define currents for each individual carrier subsystems, which
we will refer to as electron and hole quasiparticle flows, $\bb{j}_e$
and $\bb{j}_h$, respectively. The electric current is then given by
their difference, ${\bb{j}=\bb{j}_h-\bb{j}_e}$.

In external magnetic field, the system exhibits the classical Hall
effect: a voltage is generated across the system in the transverse
direction to the electric current. In a typical transport measurement,
external leads are attached to the sample in such a way, that no
current is allowed to flow in the direction of the Hall
voltage. Theoretical description of the effect is most transparent in
isotropic systems, where
${\sigma^{\alpha\beta}=\sigma_0\delta^{\alpha\beta}}$. If we
associate the $x$-axis with the electric current and the $z$-axis with
the magnetic field, then the Hall voltage is generated in the $y$
direction, while ${J_y=0}$. In two-component systems, the latter
condition leads to a field-dependent longitudinal resistivity
\cite{Weiss1954,Gant}
\be
\label{DrudeInf0}
\rho^{xx} = \frac{1}{\sigma_0}
\frac{\sigma_0^2 + \tilde\sigma_0^2  \upmu_e \upmu_h B^2}
{\sigma_0^2 + e^2(n_{0,e} - n_{0,h})^2 \upmu_e^2 \upmu_h^2  B^2},
\e
where $B$ is the magnetic field, $n_{0,e}$ and $n_{0,h}$ stand for the
equilibrium electron and hole densities, and $\upmu_e$ and $\upmu_h$
are the electron and hole mobilities. Within the standard Drude theory
\cite{Pippard,abrikos,Kittel1963}, the conductivity $\sigma_0$ can be 
expressed in terms of the quasiparticle densities and mobilities as
\[
\sigma_0= en_{0,e}\upmu_e+en_{0,h}\upmu_h, 
\]
whereas the additional parameter $\tilde\sigma_0$ is
\[
\tilde\sigma_0 = e\sqrt{\upmu_e\upmu_h(n_{0,e}^2+n_{0,h}^2) 
+  n_{0,e}n_{0,h} (\upmu_e^2+ \upmu_h^2)}.
\]
In the presence of the electron-hole symmetry, the mobilities of the
two types of carriers coincide, ${\mu_e=\mu_h=\mu}$, and the
resistivity (\ref{DrudeInf0}) simplifies to
\be
\label{DrudeInf}
\rho^{xx}=\frac{\rho_0}{e\upmu}\frac{1+(\upmu B)^2}{\rho_0^2+n_0^2(\upmu B)^2},
\e
where we have introduced quasiparticle and charge densities,
${\rho_0=n_{0,e}+n_{0,h}}$ and ${n_0=n_{e,0}-n_{h,0}}$, respectively.

The results (\ref{DrudeInf0}) and (\ref{DrudeInf}) yield a positive
magnetoresistance that is quadratic in weak magnetic fields and
saturates in classically strong fields. The two exceptions are
provided by neutral systems (${n_0=0}$, ${n_{0,e}=n_{0,h}}$), where
the quadratic magnetoresistance is non-saturating, and single-component
systems (e.g. for purely electronic transport ${n_0=\rho_0=n_{0,e}}$,
${n_{0,h}=0}$), where the longitudinal resistivity is independent of
the magnetic field \cite{Pippard,abrikos,Kittel1963}.

\begin{figure}[t]
\centerline{\includegraphics[width=0.91\columnwidth]{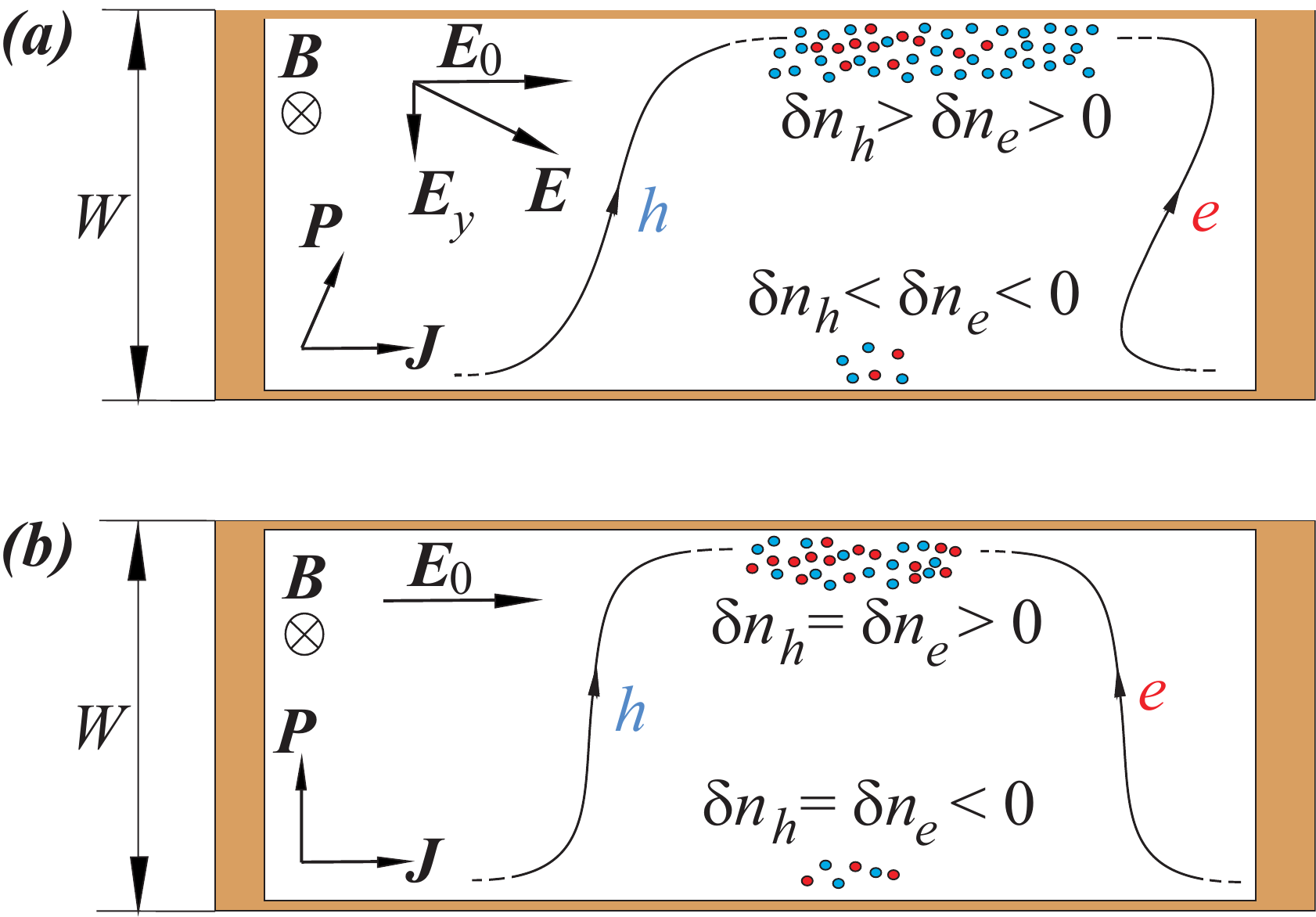}} 
\caption{Typical semiclassical trajectories for oppositely charged
  quasiparticles in two-component systems at charge neutrality. The
  two panels illustrate electron-hole asymmetric (a) and symmetric (b)
  systems. As a manifestation of the compensated Hall effect, both
  quasiparticle currents are flowing in the same direction in the bulk
  of the sample. In the symmetric sample (b), the quasiparticle flow,
  ${\bb{P}=\bb{j}_e+\bb{j}_h}$, is orthogonal to the electric
  current. In the asymmetric case (a), the longitudinal component of
  $\bb{P}$ is also finite. Such a flow leads to quasiparticle
  accumulation at the boundaries of the otherwise homogeneous
  sample. The excess quasiparticle density appears in a boundary
  region of the width of the order of the electron-hole recombination
  length. Contributions of the bulk and boundary regions to the sheet
  resistance exhibit different dependence on the magnetic field. In
  classically strong fields, the boundary region may dominate leading
  to linear magnetoresistance.}
\label{fig:qualpic}
\end{figure}

\begin{figure}[t]
\centerline{\includegraphics[width=0.95\columnwidth]{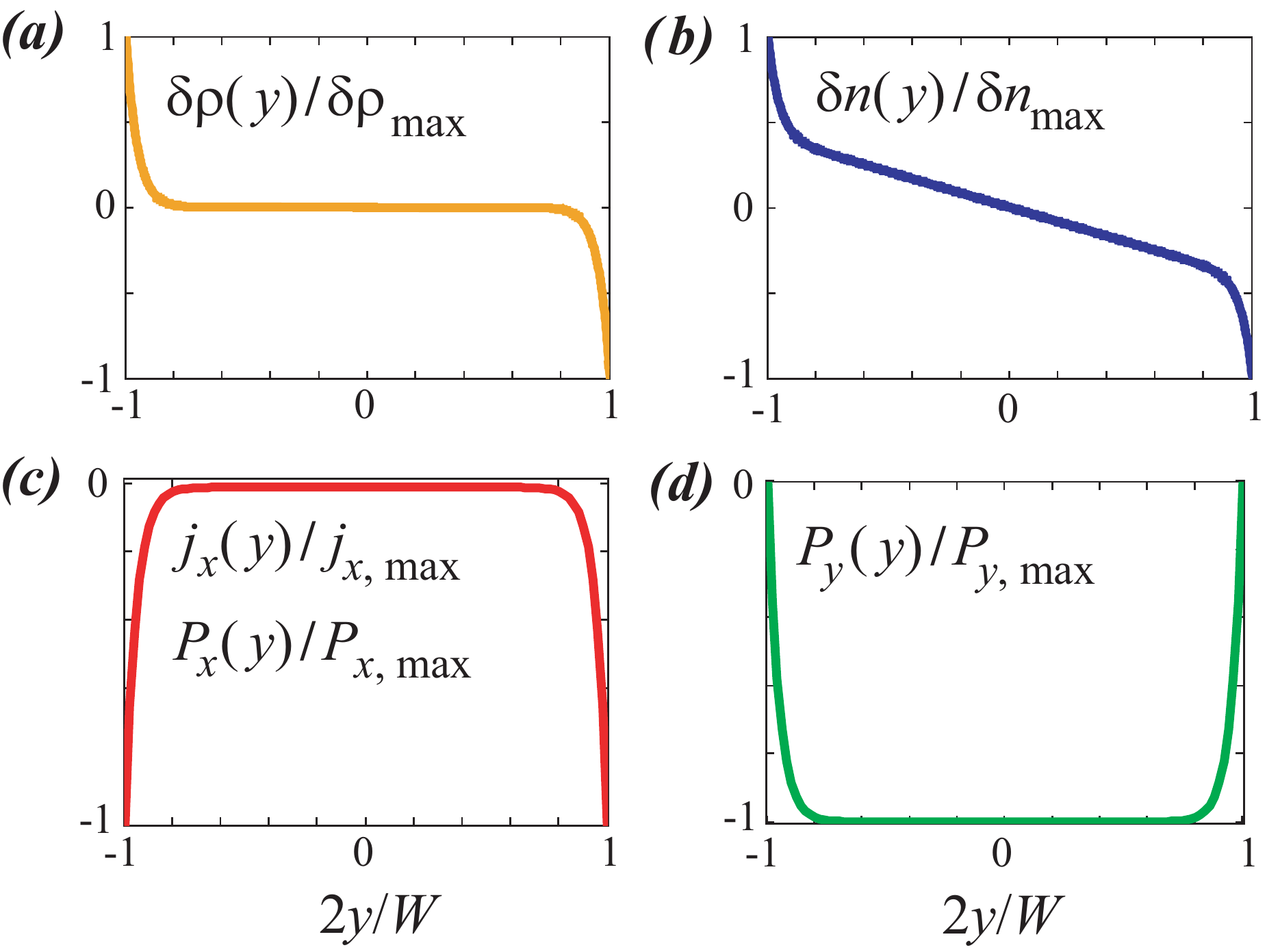}}
\caption{Lateral profiles of the quasiparticle density
  ${\delta\rho(y)}$, charge density ${\delta n(y)}$, quasiparticle
  flow ${P_{x,y}(y)}$ and electric current density ${J_x(y)}$ in a 2D
  two-component system away from charge neutrality, see
  Fig.~\ref{fig:qualpic}, panel (a), calculated within the theory
  presented in Sec.~\ref{sec:bol}. For concreteness, we chose the
  carrier parameters of a typical topological-insulator film: electron
  and hole mobilities ${\upmu_e=20\upmu_h}$,
  ${\upmu_h=1}$~m$^2/$(V$\cdot$s) and velocities ${v_e=10^6}$~m$/$s,
  ${v_h=0.5v_e}$. The sample is assumed have the width ${W=10\mu}$m,
  with the distance to the gate ${d=0.5\mu}$m and the dielectric
  constant of the surrounding insulator ${\epsilon=5}$. The carrier
  densities were calculated using a generic two-band model with the
  energy gap ${\Delta=4}$~meV at room temperature ${T=300}$~K. The
  recombination length is assumed to take the value
  ${\ell_R=0.46\mu}$m at ${B=2}$~T. All curves are normalized to the
  maxima of their absolute values.}
\label{fig:picture2}
\end{figure}

Previously \cite{Titov2013,Alekseev15}, we have pointed out an
inconsistency that appears when the above classical theory is applied
to finite-sized samples. Indeed, even partially compensated Hall
effect is accompanied by a neutral quasiparticle flow in the direction
transversal to that of the electrical current, see
Figs.~\ref{fig:qualpic} and \ref{fig:picture2}. As the quasiparticles
cannot leave the sample, this flow leads to quasiparticle accumulation
near the sample boundaries. The excess quasiparticle density is
controlled by inelastic recombination processes that are excluded from
the classical theory. The typical length scale characterizing such
processes, $\ell_R$, hereafter referred to as the recombination
length, determines the size of the boundary region with excess density
of quasiparticles. Here we consider rectangular samples with the length $L$
being the longest length scale in the system\cite{footnote},
\begin{equation}
\ell_R, \ell_R{\upmu{B}}, W\ll{L}. 
\end{equation}
The classical results are applicable if the boundary regions are small
as compared to the sample width, ${\ell_R\ll{W}}$. If, on the other
hand, $\ell_R$ is comparable with $W$, then the behavior of the system
may strongly deviate from the predictions of the classical theory.

Treating the bulk and boundary regions as parallel conductors, we
estimate the sheet resistance of the sample \cite{Alekseev15}
\begin{equation}
\label{gen0}
R_\square=\frac{W}{L}\frac{1}{R^{-1}_\textrm{bulk}
+
R^{-1}_\textrm{edge}}.
\end{equation}
In the bulk, the lateral quasiparticle flow leads to the so-called
``geometric'' magnetoresistance \cite{Lakeou,Chang2014}
\[
R_\textrm{bulk} \approx \frac{L}{W} \rho^{xx}
\quad\Rightarrow\quad
R^{-1}_\textrm{bulk} \approx \frac{W}{L}e\upmu\rho_0
\left[\frac{n_0^2}{\rho_0^2}+\frac{1}{\upmu^2B^2}\right],
\]
where we have used Eq.~(\ref{DrudeInf}) in the limit of classically
strong magnetic fields, ${\upmu{B}\gg1}$.

In the boundary regions, the quasiparticle flows are mostly directed
along the external electric field, see Figs.~\ref{fig:qualpic} and
\ref{fig:picture2}, and the geometric enhancement does not take
place. Instead, the field dependence of the edge contribution to the
sample resistance,
\[ 
R_\textrm{edge}\approx\frac{L}{\ell_{R}}\rho^{xx}(B=0),
\]
is due to the recombination length, $\ell_R$. In homogeneous samples,
the simplest estimate \cite{Titov2013,Alekseev15} yields $\ell_R$ that
is inverse proportional to $B$ in classically strong fields
\begin{equation}
\label{formula_for_l_R}
\ell_{R} = \frac{\ell_{0}}{\sqrt{1+\upmu^2 B^2}} \rightarrow 
\frac{\ell_{0}}{\upmu B},
\end{equation}
where ${\ell_{0}=2\sqrt{D\tau_{R}}}$ is the zero-field recombination
length determined by the diffusion coefficient $D$ and the
characteristic recombination time $\tau_{R}$.

The asymptotic behavior (\ref{formula_for_l_R}) of the recombination
length may be qualitatively understood as follows. In classically
strong magnetic fields, ${\upmu{B}\gg1}$, the charge carriers move
over a typical distance $R_c$ (the cyclotron radius) during a typical
diffusion time $\tau$. Since the quasiparticle life-time is determined
by the recombination processes, the overall distance covered by the
electron during the time $\tau_R$ may not exceed
${R_c\sqrt{\tau_R/\tau}\sim1/B}$, which yields the estimate for the
size of the boundary regions.

Combining the above arguments, we arrive at the following expression
for the sheet resistance (\ref{gen0}) in classically strong magnetic
fields, ${\upmu{B}\gg1}$,
\begin{equation}
\label{charge_neutr_parallel_chann}
R_\square=\frac{1}{e\rho_0\upmu}
\left[\frac{n_0^2}{\rho_0^2}+\frac{1}{\upmu^{2}B^2}+\frac{\ell_{0}}{\upmu B W}\right]^{-1}.
\end{equation}
The sheet resistance (\ref{charge_neutr_parallel_chann}) exhibits all
qualitative features of the magnetoresistance in nearly compensated
two-component systems.

In {\it wide samples}, ${W\gg\ell_{0}\upmu{B}}$, magnetotransport is
dominated by the bulk and can be described by the classical theory,
see Eqs.~(\ref{DrudeInf0}) and (\ref{DrudeInf}) and the subsequent
discussion. We consider such samples as essentially infinite.

Deviations from the classical behavior (\ref{DrudeInf0}) and
(\ref{DrudeInf}) occurs in finite-size samples of the width belonging
to the {\it intermediate} interval determined by the magnetic field,
\[
\frac{\ell_{0}}{\mu B} \ll W \ll \mu B \ell_{0}.
\]
In this case, the sheet resistance of compensated (neutral, ${n_0=0}$)
systems is linear in the magnetic field
\begin{equation}
\label{estimate}
R_\square = \frac{1}{e\rho_0}\frac{W}{\ell_{0}} B.
\end{equation}
Away from charge neutrality, LMR appears only in an intermediate range
of magnetic fields. In strong fields,
${B\gtrsim\ell_0\rho_0^2/(\upmu{W}n_0^2)}$, magnetoresistance
saturates.

In {\it narrow samples}, ${W\ll\ell_R\sim\ell_{0}/\upmu{B}}$,
recombination is ineffective and the above physical picture breaks
down. In this case, the two carrier subsystems behave as two
independent single-component systems. As a consequence, classical
magnetoresistance is absent \cite{Pippard,abrikos,Kittel1963}.

\begin{figure}[tb]
\centerline{
\includegraphics[width=0.95\columnwidth]{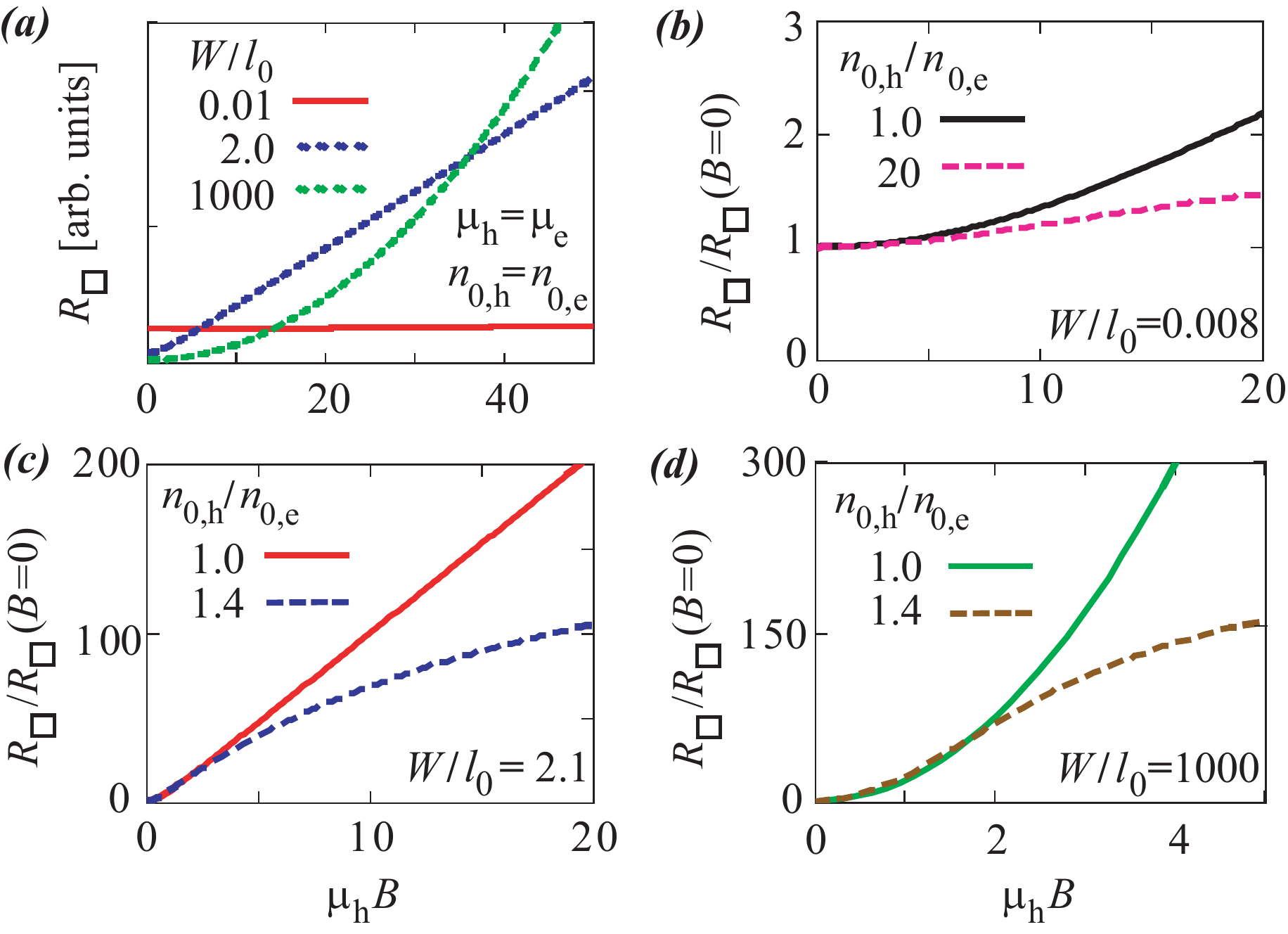}} 
\caption{(a) Sheet resistance $R_{\square}(B)$ of a 2D symmetric,
  two-component system at charge neutrality for three different values
  of the ratio of the sample width to the zero-field recombination
  length, ${W/\ell_0=0.01,2,1000}$, represented by the solid red,
  dotted blue, and dashed green lines, respectively (the curves are
  rescaled for clarity). (b), (c), (d) $R_{\square}(B)$ of a 2D
  asymmetric two-component system for three different values
  ${W/\ell_0=0.008}$ [panel (b)], ${W/\ell_0=2.1}$ [panel (b)], and
  ${W/\ell_0=1000}$ [panel (c)]. In all three plots, solid lines
  correspond to the charge neutrality point, while dashed lines show
  results away from neutrality. The curves were calculated using the
  theory presented in Sec.~\ref{sec:bol}, with the parameter values
  correspond to typical topological-insulator films (see the caption
  to Fig.~\ref{fig:picture2}), with the recombination length
  $\ell_0=4.8\mu$m.}
\label{fig:rasym}
\end{figure}

The sheet resistance (\ref{charge_neutr_parallel_chann}) is
illustrated in Fig.~\ref{fig:rasym} where it is plotted in a wide
range of classically strong magnetic fields in the above three
regimes. Panel (a) shows $R_{\square}$ for a symmetric system at
charge neutrality, while panels (b), (c), and (d) illustrate our
results for asymmetric systems at (solid curves) and away from (dashed
curves) the compensation point.

The above semiclassical mechanism of LMR in finite-size, nearly
compensated two-component system was first suggested in
Ref.~\onlinecite{Titov2013} in the context of Coulomb drag
\cite{Narozhny}. The results (\ref{gen0})-(\ref{estimate}) were
derived rigorously in graphene \cite{hydrolin} on the basis of a
microscopic transport theory. Subsequently, the macroscopic equations
derived in graphene were generalized to a generic compensated
two-component system using a phenomenological approach
\cite{Alekseev15}.

In this paper, we justify the phenomenological approach of
Ref.~\onlinecite{Alekseev15} and derive the LMR for a wide range of
systems using the Boltzmann kinetic equation. The key point that makes
our theory so general, is the simple fact that in an magnetic field
charge carriers driven through the system by the external electric
field experience a lateral drift in the direction
(${\bb{E}\times\bb{B}}$) defined by the electric and magnetic fields.
The ultimate cause of this drift is the Lorenz force that acts on all
charge carriers independently of their density, mobility, details of
the spectrum, and additional quantum numbers. The second essential
feature of our theory is the presence of the boundary leading to
accumulation of the excess quasiparticle density in the narrow regions
near the sample edges. Again, this is a completely general feature
since all samples used in laboratory (as well as all industrial
electronic devices) have a finite size. The width of the boundary
regions (and hence, the degree of macroscopic inhomogeneity in the
system, see Fig.~\ref{fig:picture2}) is controlled by the
quasiparticle recombination length. The particular dependence
(\ref{formula_for_l_R}) of $\ell_R$ on the magnetic field is crucial
for the resulting LMR, given by Eq.~(\ref{estimate}). The original
estimate \cite{Alekseev15} (\ref{formula_for_l_R}) is not universal
\cite{Vasileva16} insofar that the coefficient of the inverse
proportionality ${\ell_R\propto1/B}$ (in classically strong fields) is
system (or model) dependent. In a sense, the technical goal of the
microscopic theory presented in this paper is to calculate the field
dependence of the effective recombination length.

In our qualitative arguments, we have tacitly assumed that the energy
transfer plays no role in formation of the macroscopic inhomogeneities
of the quasiparticle currents and densities. At the microscopic level,
this means energy relaxation (and hence, thermalization) in the system
is much faster than quasiparticle recombination. As a result, the
temperatures of both carrier subsystems are uniform within the sample
(and are, in fact, identical).

The theory of Refs.~\onlinecite{Titov2013,hydrolin,Alekseev15}, as
well as the present qualitative discussion and the microscopic theory
of Sec.~\ref{sec:bol}, is focused on 2D systems.  Similar behavior can
be found also in 3D samples. In particular, if cyclotron orbits do not
remove the carriers from a plane parallel to one of the sample faces,
a linear regime similar to Eq.~(\ref{estimate}) may be observed. In
this paper, we make the first steps towards a full microscopic
understanding of magnetotransport in 3D two-component systems, see
Sec.~\ref{sec:3D}.

\section{Transport theory of 2D two-component systems}
\label{sec:bol}

In this Section we show that the linear dependence of resistivity on
the sufficiently strong magnetic field is a generic effect for
two-component systems at charge neutrality. For brevity, we employ the
natural system of units where ${\hbar=c=k_B=1}$.

The usual starting point for developing a microscopic transport theory
is the kinetic equation \cite{abrikos}. For a generic two-component
electronic system, the kinetic equation has the standard form
\be
\label{A4}
{\bb{v}}_\alpha \frac{\pa f_\alpha}{\pa \bb{r}} + e_\alpha
\lt(\bb{E}+\bb{v}_\alpha \times\bb{B} \rt) \frac{\pa f_\alpha}{\pa
  \bb{p}} = {\rm St}[f_\alpha].
\e
The semiclassical distribution functions
${f_\alpha=f_\alpha(\ep,\bb{p},\bb{r})}$ describe the positively and
negatively charged quasiparticles (``holes'' and ``electrons'',
respectively, distinguished by the index ${\alpha=e,h}$) with the
energies $\ep_\alpha(\bb{p})$ and velocities
${\bb{v}_\alpha=\pa\ep_\alpha(\bb{p})/\pa\bb{p}}$. The system is
subjected to the external electric and magnetic fields $\bb{E}$ and
$\bb{B}$.

The collision integral in the right hand side of Eq.~\eqref{A4}
comprises contributions from impurity, electron-phonon, and
electron-electron scattering. We will describe these scattering
processes by the typical time scales $\tau_{\textrm{imp}}$,
$\tau_{\textrm{ee}}$, and $\tau_{\textrm{ph}}$. The impurity and
electron-phonon scattering contribute to momentum relaxation, while
the electron-electron and electron-phonon interactions determine the
thermalization properties of the system, as well as quasiparticle
recombination. The traditional transport theory
\cite{Pippard,abrikos,Kittel1963} assumes that in the absence of
external fields the system is in equilibrium. The electric current (or
more generally, the quasiparticle flows) appears as a response to the
applied fields. Within linear response, the system experiences no
heating and remains thermalized. In this (and the following) Section
we work under the same assumptions.

Finding a general solution to the kinetic equation (\ref{A4}) is a
complicated task that is best accomplished numerically. In the special
case of Dirac fermions in graphene, the solution is facilitated by the
so-called collinear scattering singularity \cite{hydrolin}. Otherwise,
an analytical solution can be found in the two paradigmatic limiting
cases, known as the ``disorder-dominated'' and ``hydrodynamic''
regimes \cite{abrikos,Narozhny}, which can be distinguished by
comparing the scattering rates for elastic and inelastic processes:

(i) in the {\it disorder-dominated} regime, the
fastest scattering process in the system is due to potential disorder,
\be
\tau_{\rm imp} \ll \tau_{\rm ee}, \tau_{\rm ph}.
\label{A11}
\e 
Since the electron-electron scattering time is typically inverse
proportional to temperature,
\[
\tau_{\rm ee}^{-1}\propto T,
\]
the relation (\ref{A11}) implies
\[
T\tau_{\rm imp} \ll 1,
\]
which means that the motion of the charge carriers is diffusive. In
this case, most of the transport coefficients can be expressed in
terms of the diffusive constant. As a result, qualitative features of
the physical observables are independent of the microscopic details,
such as the precise form of the single-particle spectrum.

(ii) in the {\it hydrodynamic} regime, the
fastest process is due to electron-electron interaction
\be
\tau_{\rm ee} \ll \tau_{\rm imp}, \tau_{\rm ph}.
\label{A1}
\e 
Now, the relation between the temperature and the impurity scattering
time is reversed,
\[
T\tau_{\rm imp} \gg 1,
\]
so that the motion of charge carriers is ballistic. In this limit, the
system of charged quasiparticles behaves similarly to a fluid and is
described by the hydrodynamic equations. 

Remarkably, in both regimes the resistance of 2D two-component systems
close to charge neutrality exhibits linear dependence on the
orthogonal magnetic field (in sufficiently strong fields).

\subsection{Disorder-dominated regime}
\label{ddr}

\subsubsection{Symmetric, parabolic bands at charge neutrality}

We begin with the simplest case of the symmetric parabolic spectrum
with the band gap $\Delta$,
\begin{equation}
\label{A3}
\ep_{e}({\bb{p})}=\ep_{h}({\bb{p})}=\ep_{\bb{p}}=\Delta/2+ p^2/2m,
\end{equation}
where the quasiparticle velocity is proportional to the momentum
\[
\bb{v}_{\alpha}=\bb{p}_{\alpha}/m.
\]
Furthermore, we will assume the energy-independent momentum relaxation
time 
\[
\tau_h(\ep)=\tau_e(\ep)=\tau=const.
\]
At charge neutrality, the equilibrium state of the system is
described by the Fermi distribution function with the zero chemical
potential
\[
f_{\bb{p}}^{(0)}=\frac{1}{1+e^{\ep_{\bb{p}}/T}}.
\]
Since the single-particle spectrum (\ref{A3}) depends only on the 
momentum, the equilibrium quasiparticle density is given by
\be
\label{r0}
\rho_0= 2 g \int\frac{d^2\bb{p}}{(2\pi)^2}\, f^{(0)}_{\bb{p}},
\e
where $g$ is the degeneracy factor reflecting other possible
quantum numbers, such as spin, valley, etc.

External fields drive the system out of equilibrium, giving rise to
deviations of the quasiparticle densities from the equilibrium value
(\ref{r0})
\be
\label{n}
\delta n_\alpha= g \int \frac{d^2\bb{p}}{(2\pi)^2}\, f_\alpha-\frac{\rho_0}{2},
\e
and
the corresponding flow densities $\bb{j}_\alpha$:
\be
\label{j}
\bb{j}_\alpha= g \int \frac{d^2\bb{p}}{(2\pi)^2}\,\bb{v} f_\alpha.
\e
The nonequilibrium densities $\delta n_\alpha$ and currents
$\bb{j}_\alpha$ are related by the continuity equations that can be
derived by integrating the kinetic equation (\ref{A4})
\be
\label{cont-eh}
\dv{\bb{j}}_{e(h)} = -\frac{\delta n_h+\delta n_e}{2\tau_R}.
\e
Here $\tau_R$ denotes the quasiparticle recombination time. The
recombination processes typically involve electron-phonon scattering,
although in certain circumstances electron-electron \cite{Narozhny}
and three-particle \cite{hydro1} collisions may also contribute. A
calculation of the recombination time $\tau_R$ using a particular
microscopic model is beyond the scope of the present paper.

Macroscopic equations \cite{hydrolin} for the flow densities (\ref{j})
can be obtained by multiplying the kinetic equation (\ref{A4}) by the
quasiparticle velocity and summing over all single-particle states. As
a result, we find \cite{hydro1,hydrolin} 
\be
\label{A16}
\bb{\nabla}\lt[
g \!\int\! \frac{d^2\bb{p}}{(2\pi)^2}\,\frac{v^2}{2}f_\alpha\rt]
-\frac{e_\alpha \bb{E} \rho_0}{2m}-\bb{j}_\alpha \!\times\! \bb{\omega}_\alpha
=-\frac{\bb{j}_\alpha}{\tau},
\e
where ${\bb{\omega}}_h\!\!=\!-{\bb{\omega}}_e\!\!=\!{\bb{\omega}_c}$ are
the carrier cyclotron frequencies ${\bb{\omega}_c\!=\!e\bb{B}/m}$.

Comparing the integral in Eq.~(\ref{A16}) with the flow density
(\ref{j}), we find it natural to split the distribution functions
$f_\alpha$ into the ``isotropic'' and ``anisotropic'' parts,
\be
\label{A12}
f_\alpha =f_\alpha^{(i)}(\ep) +f_\alpha^{(a)}(\ep,\bb{e}_{\bb{p}}).
\e 
The isotropic term depends only on the quasiparticle energy and hence
does not contribute to the currents (\ref{j}). On the contrary, the
anisotropic term is an odd function of the momentum. It is this part
of the distribution function that determines the currents (\ref{j}),
but at the same time, it does not contribute to the integral in
Eq.~(\ref{A16}).

Within linear response, deviations of the isotropic function
$f_\alpha^{(i)}(\ep)$ from the equilibrium distribution
$f^{(0)}_{\bb{p}}$ can either reflect deviations of the local
electronic temperature from the equilibrium value determined by the
lattice, or the local fluctuations of the chemical potential
${\delta\mu_{\alpha}(\bb{r})}$. 

Thermalization between the electronic system and the lattice is
achieved by means of electron-phonon coupling. While the same coupling
is also responsible for quasiparticle recombination, the latter is a
much slower process and does not affect the local temperature.
Relegating a more detailed discussion of this issue to a future
publication, hereafter we assume that the relation
\[
\tau_{\rm ph} \ll \tau_R
\]
allows us to neglect local temperature fluctuations
\[
\delta T(\mathbf{r})=0.
\]
As a result, the isotropic part of the distribution function may only
depend on the local fluctuations of the chemical potential
\be
\label{delta-f}
f_\alpha^i=f^{(0)}_{\bb{p}}+\frac{\pa f^{(0)}_{\bb{p}}}{\pa \ep} \delta \mu_\alpha(\bb{r}).
\e
This implies the proportionality between the local density
fluctuations (\ref{n}) and ${\delta\mu_\alpha(\bb{r})}$:
\be
\label{nabla-n}
\delta n_\alpha=\nu_0 \delta \mu_\alpha,
\e
where (cf. Ref.~\onlinecite{hydrolin})
\be
\label{average}
\nu_0 = \la 1 \ra, \qquad
\la \cdots \ra = -g\!\!\int\limits_{\Delta/2}^{\infty}\!\!d\ep
\nu(\ep)\frac{\pa f^{(0)}}{\pa \ep}(\cdots),
\e
with $\nu(\ep)$ being the density of states [$\nu_0$ has dimensions of
  $\nu(\ep)$].

Since the equilibrium distribution $f^{(0)}_{\bb{p}}$ is
independent of $\bb{r}$, we can express the integral in
Eq.~(\ref{A16}) as
\[
g\!\int\!\frac{d^2\bb{p}}{(2\pi)^2}\,\frac{v^2}{2} f_\alpha
=\frac{\lt\la v^2\rt\ra}{2} \delta \mu _{\alpha}
=\frac{\lt\la v^2\rt\ra}{2\nu_0}\delta n_\alpha,
\]
and introduce the diffusion coefficient in Eq.~\eqref{A16} 
\be
\label{currents-eh}
D \bb{\nabla} \delta n_\alpha - e_\alpha \bb{E} \rho_0\tau/(2m)
-\bb{j}_\alpha\!\times\!\bb{\omega}_\alpha \tau =-\bb{j}_\alpha.
\e
The diffusion coefficient is the same for the electrons and holes:
\be
D=\la v^2\ra\tau/(2\nu_0).
\e
At charge neutrality the averages in the expression for the diffusion
coefficient can be evaluated analytically:
\be
\label{D}
D(\mu\!=\!0)=\frac{T\tau}{m} \lt(1+e^{\Delta/2T}\rt)\ln\lt(1+e^{-\Delta/2T}\rt).
\e

The macroscopic equations (\ref{cont-eh}) and (\ref{currents-eh})
allow us to find transport coefficients of the system, as well as the
carrier density and current profiles. These equations are
semiclassical in the sense that the effects of quantum interference
\cite{ZNA} and Landau quantization
\cite{abrikos,Abrikosov1969,Abrikosov1998,Abrikosov2000} are
neglected.

In this paper we are interested in solving the macroscopic transport
equations (\ref{cont-eh}) and (\ref{currents-eh}) in confined
geometries (in fact, that is why we have considered the nonuniform
distributions). For simplicity, we consider a rectangular sample with
the length that is much larger than the width ${L\gg{W}}$, as well as
any correlation length in the system. In this case, all physical
quantities depend only on the transversal coordinate $y$
(${-W/2<y<W/2}$). If no contacts are attached to the side edges of the
sample, the quasiparticle flows have to vanish at the edges
\begin{equation}
\label{bc}
j^y_\alpha(y=\pm W/2)=0.
\end{equation}

Combining the carrier densities (\ref{n}) into the charge density,
${\delta n\!=\!\delta n_e\!-\!\delta n_h}$, and total quasiparticle
density ${\delta\rho\!=\!\delta n_e\!+\!\delta n_h}$, and introducing
the corresponding currents, ${\bb{j}\!=\!\bb{j}_h\!-\!\bb{j}_e}$ and
${\bb{P}\!=\!\bb{j}_e\!+\!\bb{j}_h}$, we may represent the macroscopic
equations (\ref{cont-eh}) and (\ref{currents-eh}) in the form
\cite{Alekseev15}
\beml
\label{basic2}
\beq
\label{n2a}
&&
D\bb{\nabla}\delta \rho +\bb{P}- \bb{j}\!\times\!\bb{\omega}_{c}\tau=0,
\\
&&
\nonumber\\
\label{n3a}
&&
D\bb{\nabla}\delta n +\bb{j}-e\bb{E} \rho_0\tau/m- \bb{P}\!\times\!\bb{\omega}_{c}\tau=0,
\\
&&
\nonumber\\
\label{div2b}
&&
\dv \bb{P}=-\delta \rho/\tau_R, \qquad \dv\bb{j} =0.
\eq
\eml

Looking for solutions independent of the $x$ coordinate and keeping in
mind the hard-wall boundary conditions (\ref{bc}), we find 
\[
\bb{P} = P(y) \bb{e}_y, \qquad \bb{j} = j(y)\bb{e}_x,
\qquad \delta n=0.
\]
Moreover, we note that the equations (\ref{basic2}) preserve the
direction of the applied electric field if choose it to be 
\[
\bb{E}=E_0\bb{e}_x.
\]
Then we can use Eq.~(\ref{div2b}) to exclude the quasiparticle density
and simplify Eqs.~(\ref{n2a}) and (\ref{n3a}) as
\beml
\label{basic3}
\beq
\label{3a}
&&
-D\tau_R\;\pa^2 P/\pa y^2 +P(y) + \omega_c \tau j(y) =0,
\\
&&
\nonumber\\
&&
j(y)=j_0+\omega_c\tau P(y),
\label{3b}
\eq
\eml
where ${j_0=e\tau\rho_0E_0/m}$ is the electric current in the absence of
magnetic field.

The second-order differential equation (\ref{3a}) with the hard-wall
boundary conditions (\ref{bc}) admits the solution \cite{Alekseev15}
\be
\label{profile}
P(y)=j_0 \frac{\omega_c \tau}{1+\omega_c^2\tau^2}
\lt(\frac{\cosh(2y/\ell_R)}{\cosh(W/\ell_R)}-1\rt),
\e
where the quasiparticle recombination length in magnetic field is
\[
\ell_R=\ell_0/\sqrt{1+\omega_c^2\tau^2}, \qquad
\ell_0=2\sqrt{D\tau_R}.
\]

The quasiparticle current (\ref{profile}) and the corresponding
electric current $\bb{j}(y)$ are illustrated in
Fig.~\ref{fig:qualpic}. The nonuniform nature of the currents does not
allow for establishing a meaningful resistivity in our system. Instead,
we may define the sheet resistance \cite{Alekseev15}
\be
\label{Rsquare}
R_\square=E_0/\overline{J},\qquad \overline{J}
=
\frac{e}{W}\!\int\limits_{-W/2}^{W/2}\! j(y) dy.
\e
The resulting value of $R_\square$ is given by
\be
\label{final0}
R_\square\!=\!\frac{m}{e^2\tau\rho_0}
\frac{1\!+\!\omega_c^2\tau^2}{1\!+\!\omega_c^2\tau^2 F(W/\ell_R)},
\;\;
F(x)\!=\!\frac{\tanh(x)}{x}.
\e

The sheet resistance (\ref{final0}) was previously obtained in
Ref.~\onlinecite{Alekseev15} using a phenomenological
approach. Depending on the sample width $W$, recombination length
$\ell_0$, and magnetic field, one may identify three types of
asymptotic behavior \cite{Alekseev15}:

(i) in wide samples, ${W\!\gg\!(\omega_c\tau)^2\ell_{R}}$, the
resistance (\ref{final0}) is a non-saturating, quadratic function of
the $B$ field \cite{Weiss1954}
\begin{subequations}
\label{cases}
\be
\label{r1}
R_\square=\frac{m}{e^2\tau\rho_0}\lt(1+\omega_c^2 \tau^2\rt).
\e
The resistance (\ref{r1}) exhibits geometric enhancement that is a
consequence of the compensated hall effect, where the Hall voltage is
absent despite the tilt of the carrier trajectories.

(ii) in narrow samples, ${W\!\ll\!\ell_{R}}$, quasiparticle
recombination is ineffective, all currents flow along the $x$-axis,
and hence the geometric enhancement factor is absent
\be
R_\square=\frac{m}{e^2\tau\rho_0}.
\e

(iii) samples of intermediate width,
${\ell_{R}\!\ll\!W\!\ll\!\omega_c^2\tau^2\ell_{R}}$, in classically
strong magnetic fields, ${\omega_c\tau\!\gg\!1}$,
exhibit a linear behavior \cite{Alekseev15,hydrolin}
\be
\label{linear1}
R_\square=\frac{m}{e^2\tau\rho_0} \frac{W}{\ell_{R}},
\e
\end{subequations}
shown in Eq.~(\ref{estimate}) above (note that
${\omega_c\tau\!=\!\upmu{B}}$).

The results of this section provide the microscopic justification to
the phenomenological approach of Ref.~\onlinecite{Alekseev15}. Similar
results were previously obtained for monolayer graphene
\cite{hydrolin}. In the following sections we generalize our theory to
the case of arbitrary quasiparticle spectrum and prove that LMR
in classically strong fields is a generic feature of compensated,
two-component systems.

\subsubsection{Symmetric bands with arbitrary spectrum}

In this section, we generalize our kinetic theory to the case of the
arbitrary quasiparticle spectrum, $\ep(\bb{p})$, and energy-dependent
momentum relaxation time, $\tau(\ep)$. For simplicity, we only
consider rotationally invariant spectra, ${\ep(\bb{p})\!=\!\ep_p}$,
${p\!=\!|\bb{p}|}$. The cyclotron frequency is now also
energy-dependent
\be
\label{A14}
\bb{\omega}_h=-\bb{\omega}_e= \bb{\omega}_c,\qquad  
\bb{\omega_c}(\ep)=e\bb{B}\,v/p,
\e
while the velocity and momentum are given by the usual relations
\be
v(\ep)=\lt| \frac{\pa \ep_p }{\pa p}   \rt|,
\quad
p=p(\ep), \quad \ep_{ p (\ep) }=\ep.
\e

The energy dependence of the velocity and momentum relaxation time
makes the derivation of the macroscopic transport equations rather
tedious.  Instead, we use the kinetic equation \eqref{A4} to relate
the two parts of the distribution function \eqref{A12}. The
anisotropic part of the kinetic equation reads
\be
\label{kin1}
\bb{v} \bb{\nabla} f_\alpha^{(i)}
 + e_\alpha \bb{E}\bb{v} \frac{\pa f_\alpha^{(i)}}{\pa \ep}
  +\omega_\alpha(\ep) \frac{\pa f_\alpha^{(a)}}{\pa \varphi} =
- \frac{f_\alpha^{(a)}}{\tau(\ep)},
\e
where the angle $\varphi$ describes direction of the velocity.
Solving Eq.~(\ref{kin1}) for $f_\alpha^{(a)}$, we find
\be
\label{A13}
f_\alpha^{(a)} = \s_{k,l} v^k \tau^{kl}_\alpha \lt(-\frac{\pa}{\pa x^l} 
+ e_\alpha E^l \frac{\pa}{\pa \ep}\rt)f_\alpha^{(i)},
\e
where the indices ${k,l\!=\!x,y}$ indicate the 2D vector
components. The tensor $\tau^{kl}_\alpha$ is given by
\be
\hat{\tau}_\alpha=\frac{\tau(\epsilon)}{1+\omega_c^2(\ep)\tau^2(\ep)}
\bpm
1 & \omega_\alpha(\ep) \tau(\ep)\\
-\omega_\alpha(\ep)\tau(\ep)  & 1
\epm.
\e

Now we can use Eq.~(\ref{kin1}) to express the carrier flow densities
(\ref{j}) in terms of the isotropic part of the distribution function.
Instead of the direct momentum integration, we now evaluate the
currents (\ref{j}) in two steps. Firstly, we average over the
direction of the velocity. This yields the energy-dependent currents
\be
\label{jE}
{j}^k_\alpha(\ep) = D^{kl}_\alpha (\ep, B) 
\lt(-{\nabla}^l+e_\alpha {E}^l \frac{\pa}{\pa \ep}\rt) f_\alpha^{(i)},
\e
where ${\hat{D}_\alpha(\ep,B)=v^2\hat{\tau}_\alpha/2}$. Secondly, we
integrate over the energy using the explicit form \eqref{delta-f} of
the distribution function. The expression \eqref{delta-f} is still
valid, since none of the assumptions of the previous section relied on
the particular shape of the quasiparticle spectrum. Substituting
Eq.~\eqref{delta-f} into Eq.~(\ref{jE}) we find
\be
\label{A15}
j^k_\alpha(\ep) = D_\alpha^{kl} (\ep,B) 
\lt[ \nabla^l\delta \mu_\alpha(\bb{r})+e_\alpha E^l\rt] \frac{\pa  f^{(0)}}{\pa \ep}.
\e
Integrating Eq.~(\ref{A15}) over the energy, we obtain 
\be
\label{j1}
j^k_\alpha = D^{kl}_\alpha(B) \lt( - \nabla^l \delta n_\alpha +e_\alpha\nu_0 E^l \rt),
\e
with the averaged ``diffusion tensor'' is
\beml
\label{D-ave}
\be
\hat{D}_{e(h)}(B)=\frac{1}{\nu_0} \la \hat{D}_\alpha(\ep, B) \ra =
\bpm
D_{xx} & \pm D_{xy}
\\
\mp D_{xy} & D_{xx}
\epm.
\e
The individual matrix elements of $\hat{D}_{e(h)}(B)$ are given by
\beq
&&
D_{xx} = \frac{1}{\nu_0}
\lt\la \frac{v^2}{2} \frac{\tau(\ep)}{1+\omega_c^2(\ep) \tau^2(\ep)}\rt\ra,
\\
&&
\nonumber\\
&&
D_{xy} = \frac{1}{\nu_0}
\lt\la \frac{v^2}{2} \frac{\omega_c(\ep)\tau^2(\ep)}{1+\omega_c^2(\ep) \tau^2(\ep)}\rt\ra.
\eq
\eml
For the energy-independent $\tau$ and $\omega_c$ the matrix $\hat{D}_{e(h)}(B)$
simplifies to 
\be
\label{D-DB}
\hat{D}_\alpha(B)=\frac{D}{1 +\omega_c^2 \tau^2}
\bpm
1 & \omega_\alpha \tau\\
-\omega_\alpha \tau  & 1
\epm,
\e
where $D$ is given by Eq.~\eqref{D}.

The expression (\ref{j1}) generalizes the above macroscopic equation
\eqref{currents-eh} for the case of an arbitrary quasiparticle
spectrum and energy-dependent momentum relaxation rate [for the
  parabolic spectrum, we recover Eq.~\eqref{currents-eh} with the help
  of the identity ${\la{v^2/2}\ra=n_0/m}$]. The corresponding
continuity equations are still given by Eq.~(\ref{cont-eh}), where
$\tau_R$ now stands for the mean value of the recombination
time. Again, in this paper we do not study microscopic details of the
recombination processes and, in particular, the energy dependence of
the recombination rate.

At charge neutrality, the densities of electrons and holes coincide,
${\delta n_h=\delta n_e=\delta\rho/2}$. Similarly to the case of the
parabolic spectrum, the hard-wall boundary conditions (\ref{bc})
ensure that the electric field does not deviate from its
direction along the the $x$-axis, ${\bb{E}=E_0\bb{e}_x}$. This allows
us to re-write Eq.~(\ref{j1}) in the form
\beml
\label{jxjy}
\beq
\label{jx}
&&
j_h^x=-j_e^x=
e \nu_0D_{xx} E_0 + \frac{1}{2}D_{xy}\frac{\pa \delta \rho}{\pa y},
\\
&&
\nonumber\\
&&
j_h^y=j_e^y=
e \nu_0D_{xy} E_0  -\frac{1}{2}D_{xx}\frac{\pa \delta\rho}{\pa y}.
\label{jy}
\eq
\eml
Combining the currents (\ref{jxjy}) with the continuity equation
(\ref{cont-eh}), we find a second-order differential equation for
$\delta\rho$
\be
\label{ddn}
\frac{\pa^2 \delta \rho}{\pa y^2} =
\frac{4\delta\rho}{\ell_R^2},\qquad \ell_R = 2\sqrt{D_{xx} \tau_R}.
\e 

The equations (\ref{jxjy}) and (\ref{ddn}) are completely equivalent
to Eqs.~(\ref{basic3}). The only difference is the precise definition
of the diffusion coefficients. Hence, it is not surprising that the
solution to Eq.~(\ref{ddn}) with the hard-wall boundary conditions
(\ref{bc}) is similar to Eq.~(\ref{profile})
\be
\label{n(y)}
\delta\rho  = -e\nu_0E_0\ell_R
\frac{D_{xy}}{D_{xx}} \frac{\sinh(2y/\ell_R)}{\cosh(W/\ell_R)}.
\e
Finally, we use the solution (\ref{n(y)}) and Eqs.~(\ref{jxjy}) to
find the averaged electric current and
sheet resistance (\ref{Rsquare})
\be
\label{J}
R_\square = \frac{1}{2 e^2} 
\lt(D_{xx}+\frac{D_{xy}^2}{D_{xx}}F(W/\ell_R)\rt)^{-1}.
\e

Qualitatively, the result (\ref{J}) is similar to Eq.~(\ref{final0}),
see also Fig.~\ref{fig:r0rh}. Most importantly, the dependence of
$R_\square$ on the magnetic field and sample geometry is given by the
same function $F(W/\ell_R)$. Therefore, we can identify the same three
types of behavior as in Eqs.~(\ref{cases}).

(i) in the limit of a {\it wide sample} the contribution of the
function $F(W/\ell_R)$ may be neglected. The resulting
magnetoresistance is quadratic and unsaturating,
${R_\square\sim1/D_{xx}}$.

(ii) the limit of a {\it narrow sample} corresponds to the
approximation ${F(W/\ell_R)\approx1}$. In this case, the sheet
resistance (\ref{J}) is not strictly speaking a constant, but exhibits
weak, quickly saturating dependence on the magnetic field,
${R_\square\sim{D}_{xx}/(D_{xx}^2+D_{xy}^2)}$.

(iii) the limit of an {\it intermediate sample size} exists in
classically strong magnetic fields, where we may approximate
${F(x)\approx1/x}$ and neglect the field-independent term in
Eq.~(\ref{J}). This leads to the linear magnetoresistance similar to
Eq.~(\ref{linear1}). The parameter range for this regime is similar to
that of the previous section:
${\ell_{R}\!\ll\!W\!\ll\!\ell_{R}D_{xy}^2}/D_{xx}^2$. The resulting
resistance is 
\begin{equation}
\label{lmr2}
R_\square = \frac{1}{2 e^2} 
\frac{D_{xx}}{D_{xy}^2}\frac{W}{\ell_R}.
\end{equation}
The result (\ref{lmr2}) may be simplified if we formally assume the
limit ${B\to\infty}$. Then the elements of the diffusion matrix are
\[
D_{xx}= \frac{\la p^2/\tau\ra}{2\nu_0 e^2B^2},
\quad
D_{xy}= \frac{\lt\la vp\rt\ra}{2\nu_0eB}.
\]
The recombination length is inverse proportional to the magnetic field
\[
\ell_R  = \frac{1}{e B} \sqrt{\frac{2\tau_R}{\nu_0}\lt\la\frac{p^2}{\tau}\rt\ra},
\]
and hence the resistance is linear in the $B$-field
\be
\label{r2b}
R_\square({B\to\infty})=\frac{B}{e} 
\sqrt{\frac{\nu_0 \la p^2/\tau\ra}{2\tau_R}}\frac{\nu_0 W}{\la vp\ra^2}.
\e

\subsubsection{Asymmetric bands}

Now we discuss a generic two-component system without electron-hole
symmetry. For simplicity, we will consider the parabolic spectra (as
we have seen above, changing the shape of the quasiparticle spectrum
does not lead to qualitatively new physics)
\be
\label{psp}
\ep_\alpha(\bb{p})=\Delta/2+p^2/2 m_\alpha.
\e
In addition, the system may be doped away from charge neutrality, i.e.
the equilibrium chemical potential may be shifted from the middle of
the band gap. Nevertheless, we may repeat the derivation of the
continuity equations (\ref{cont-eh}) and macroscopic equations
(\ref{currents-eh}) and arrive at the following description of the 
system
\begin{eqnarray}
&&
D_{ \alpha} \bb{\nabla} \delta n_\alpha
 - e_\alpha \bb{E} n_{0,\alpha}\tau_\alpha /m_\alpha
- \bb{j}_\alpha\!\times\! \bb{\omega}_\alpha \tau_\alpha
 =- \bb{j}_\alpha,
\nonumber\\
&&
\nonumber\\
&&
\dv \bb{j}_\alpha=- (\Gamma_{e} \delta n_{e}+\Gamma_{h} \delta n_{h} ) /2.
\label{basic}
\end{eqnarray}
The electrons and holes are described by their respective densities
${\delta n_\alpha(\bb{r})=n_\alpha(\bb{r})-n_{0,\alpha}}$, masses
$m_\alpha$, momentum relaxation times $\tau_\alpha$, cyclotron
frequencies $\bb{\omega}_\alpha={e_\alpha\bb{B}}/{m_\alpha c}$, and
diffusion coefficients
\be
\label{D_0}
D_\alpha=\la v^2\ra_\alpha\tau_\alpha/(2\nu_{0,\alpha}).
\e
Here the averaging over energies is similar to Eq.~(\ref{average}),
but with the different equilibrium distribution functions for
electrons and holes, $f_\alpha^{(0)}$:
\be
\label{average1}
\langle \cdots \rangle_\alpha
=-\int\limits_{\Delta/2}^{\infty} \!\!d\ep \nu_\alpha(\ep)  
\frac{\pa f_\alpha^{(0)}}{\pa \ep}(\cdots),
\e
where $\nu_\alpha(\ep)$ is the corresponding density of states.

The recombination rates $\Gamma_\alpha$ are generally different for
electrons and holes and may be approximated as
\be
\label{Gam_e_h}
\Gamma_e= 2\gamma n_{0,h}, \qquad \Gamma_h= 2\gamma n_{0,e} ,
\e
where the coefficient $\gamma$ is the function of $T$ and
depends on a particular model of electron-hole recombination.

In the absence of the electron-hole symmetry, the classical Hall
effect is no longer completely compensated and the Hall voltage is
formed. The corresponding lateral component of the electric field can
be related to the nonuniform charge density across the sample. In
principle this can be done by solving the Poisson equation with the
sample-specific boundary conditions. This electrostatic problem can be
rather complicated and may admit only numerical solutions. While one
may have to solve the electrostatic problem to describe the behavior
of any particular sample quantitatively, qualitative physics is
independent of such complications. Here, we will consider the simplest
case of a gated sample. If the distance between the 2D electron system
and the gate electrode is much smaller than any typical length scale
describing inhomogeneity of the charge density and carrier flows, then
the system is in the {\it strong screening} limit, where the electric
field is related to the charge density as \cite{Alekseev15}
\be
\label{electro}
\bb{E}=E_0\bb{e}_{x}-\frac{e}{C}\frac{\pa \delta n}{\pa y}\bb{e}_{y},
\e
where $E_0$ is the external field,  $C=\epsilon/4\pi d$ is the
gate-to-channel capacitance per unit area, $d$ is the the distance to the
gate, $\epsilon $ is dielectric constant,
and ${\delta n = \delta n_h - \delta n_e}$ is the charge density.

The macroscopic equations (\ref{basic}) are linear differential
equations that can be solved similarly to the above case of the
symmetric bands. Before presenting the general solution, we
discuss two particular limiting cases, (i) the Boltzmann limit away
from charge neutrality, and (ii) the fast Maxwell relaxation.

\subsubsection{Boltzmann limit}

First, we consider the low-temperature (Boltzmann) limit,
${T\ll\Delta}$, where the effective number of charge carriers in both
bands is small. In the simplest case, the carriers have the same mass,
${m_\alpha=m}$, and momentum relaxation time,
${\tau_\alpha=\tau=const}$. Consequently, the two cyclotron
frequencies also coincide, ${\omega_h=-\omega_e=\omega_c}$. The
electron-hole symmetry is broken by the non-zero chemical potential,
${\mu_{0,e}=-\mu_{0,h}=\mu}$. The above parameters can be combined into
the ``Drude conductivities'' of the electrons and holes
\begin{equation}
\sigma_{e(h)}=e^2n_{0,e(h)} \tau/m.
\end{equation}

In this limit, the equilibrium distribution functions have the simple
form
\begin{equation}
f^{(0)}_{e(h)}=\exp\left(-\frac{\ep+\Delta/2\mp\mu}{T}\right),
\end{equation}
allowing for the explicit expressions for the equilibrium carrier
concentrations (with ${\nu=gm/2\pi}$ being the density of states for
the 2D parabolic spectrum)
\begin{equation}
n_{0,e(h)}= \nu T \exp \left(-\frac{\Delta/2\mp\mu}{T}\right).
\end{equation}
Furthermore, with exponential accuracy the diffusion coefficients
(\ref{D_0}) can be approximated by
\begin{equation}
D_\alpha =D=T\tau/m.
\end{equation}

The above simplifications allow for a straightforward solution of the
macroscopic equations (\ref{basic}). The averaged sheet resistance
(\ref{Rsquare}) is given by
\be
\label{RBolt}
R_\square=\frac{1}{\sigma_e+\sigma_h}
\frac{1+\omega_c^2\tau^2}{1+\omega_c^2\tau^2\lt[\xi +(1-\xi)F(W/\ell_R)\rt]},
\e
where $\xi=n^2_0/\rho^2_0$ and the magnetic-field dependent
recombination length $\ell_R$ is 
\be
\ell_R=2\sqrt{\frac{2eD}{(\Gamma_e+\Gamma_h)(1+\omega_c^2\tau^2)}}.
\e
At charge neutrality, ${\xi=0}$, we recover the previous result
(\ref{final0}). The magnetoresistance $R_\square(B)$ is shown in
Fig.~\ref{fig:rasym} for several values of $\xi$.

Since the system is doped away from charge neutrality, the classical
Hall effect is no longer fully compensated. This can be seen in the
solution to the equations (\ref{basic}), where the electric field
acquires a constant component in the lateral direction
\be
\label{Eyy}
E_y= -\frac{\omega_c\tau E_0(\sigma_e-\sigma_h)}{\sigma_h+\sigma_e+DC},
\e
leading to the nonzero Hall voltage, ${V_H=E_yW}$. The corresponding
Hall sheet resistance
\[
R_\square^{H}=E_y/\overline{J}=R_\square E_y/E_0,
\]
is given by
\beq
&&
R_\square^{H}=-\frac{\omega_c\tau}{\sigma_e+\sigma_h+DC}
\frac{\sigma_e-\sigma_h}{\sigma_e+\sigma_h}
\\
&&
\nonumber\\
&&
\qquad\qquad\qquad
\times
\frac{1+\omega_c^2\tau^2}{1+\omega_c^2\tau^2\lt[\xi +(1-\xi)F(W/\ell_R)\rt]}.
\nonumber
\eq

\subsubsection{Fast Maxwell relaxation}

A more general situation with unequal carrier masses and momentum
relaxation times also allows for a simple solution under the
assumption of fast Maxwell relaxation,
\[
C\ll m_\alpha e^2.
\]
In this limit, charge fluctuations in the two-component system relax
much faster than the usual diffusion.

Formally taking the limit ${C\to{0}}$ in Eqs.~(\ref{basic}) and
(\ref{electro}), we recover the balance between the nonequilibrium
density fluctuations of the electrons and holes 
\be
\label{maxwell}
\delta n_e = \delta n_h =\delta \rho/2.
\e
Note, that this does not imply charge neutrality, since these
fluctuations occur on the background of nonzero equilibrium charge
density $n_0$.

Now, we can express the quasiparticle flows (\ref{j}) in terms of the
density perturbation (\ref{maxwell}) and electric field
\be
\label{jjmatrix}
j^k_\alpha =
\lt(
\frac{ e_\alpha E^l n_{0,\alpha}\tau_\alpha}{m_\alpha}
- \frac{ D_\alpha \nabla^l \delta \rho}{2}  \rt) \tau^{lk},
\e
where
\be
\hat{\tau}=\frac{1}{1 +\omega_c^2 \tau_\alpha^2}
\bpm
1 & \omega_\alpha \tau_\alpha \\
-\omega_\alpha\tau_\alpha  & 1
\epm.
\e
Here the cyclotron frequency, ${\omega_\alpha=e_\alpha B/m_\alpha}$, has
the opposite sign for electrons and holes.

The hard-wall boundary conditions (\ref{bc}) imply the equality
${j^y_e=j^y_h}$. Excluding the $y$-component of the electric field
from Eqs.~(\ref{jjmatrix}), we can express the currents $j^y_\alpha$
in terms of the quasiparticle density $\delta\rho$. This allows us to
re-write the continuity equations in the form of the second-order
differential equation on ${\delta\rho(y)}$, same as Eq.~(\ref{ddn}),
which we reproduce here for convenience,
\be
\label{drho}
\frac{d^2 \delta\rho}{dy^2}
=\frac{4\delta\rho}{\ell_R^2}.
\e
In contrast to Eq.~(\ref{ddn}), the effective recombination length
is now given by
\be
\label{kappa1}
\ell_R =
2\sqrt{\frac{\sigma_e^{xx} D_h^{xx} + \sigma_h^{xx} D_e^{xx}}
{(\Gamma_e+\Gamma_h) (\sigma_e^{xx}+\sigma_h^{xx})}},
\e
where the two-component quasiparticle system is characterized by
the field-dependent conductivity matrix  
\beml
\label{sdb}
\be
\hat{\sigma}_\alpha =
\bpm
\sigma_\alpha^{xx} & \sigma_\alpha^{xy}
\\
\sigma_\alpha^{yx}  & \sigma_\alpha^{xx}
\epm
=\frac{e^2n_\alpha^0\tau_\alpha}{m_\alpha} \hat{\tau},
\e
and the field-dependent diffusion matrix
\be
\hat{D}_\alpha(B) =
\bpm
D_\alpha^{xx} & D_\alpha^{xy}
\\
D_\alpha^{yx}  & D_\alpha^{xx}
\epm
=D_\alpha \hat{\tau}.
\e
\eml

The solution to Eq.~\eqref{drho}, which satisfies the boundary
conditions, differs from the previous result (\ref{n(y)}) by the
normalization factor that is dictated by the relation (\ref{jjmatrix})
between the density and quasiparticle flows. In the present case we
find
\be
\delta\rho=-E_0\ell_R
\frac{\sigma_e^{xx}|\sigma_h^{xy}|+|\sigma_e^{xy}|\sigma_h^{xx}}
{\sigma_e^{xx}D_h^{xx}+\sigma_h^{xx}D_e^{xx}}\frac{\sinh(2y/\ell_R)}
{\cosh(W/\ell_R)}.
\label{drho1}
\e
Substituting Eq.~\eqref{drho1} into Eq.~\eqref{jjmatrix}, we express
the inverse sheet resistance in the form
\be
\label{f1}
R^{-1}_\square=\lt(\rho_{\infty}^{xx}\rt)^{-1}+A F(W/\ell_R),
\e
where $\rho_{\infty}^{xx}$ is the resistivity of an infinitely large
system
\be
\label{f2}
\hat \rho_\infty =
\lt(\hat \sigma_e+\hat \sigma_h\rt)^{-1}.
\e
and
\be
\label{A}
A=\frac{(\sigma_e^{xx}|\sigma_h^{xy}|+|\sigma_e^{xy}|\sigma_h^{xx})^2}
{(\sigma_e^{xx}+\sigma_h^{xx})\sigma_e^{xx}\sigma_h^{xx}}.
\e

The result (\ref{f1}), as well as the result of the previous section,
Eq.~(\ref{RBolt}), has the same functional dependence on the magnetic
field as our previous result, Eq.~(\ref{J}). Therefore, also in the
present case we can identify the three limiting cases of the wide,
narrow, and intermediate-sized samples. In the latter case, we recover
the linear dependence on the magnetic field. However, in contrast to
the case of the neutral system, described by Eq.~(\ref{J}), the system
away from charge neutrality exhibits saturation of the linear
behavior. For illustration, we consider the formal limit
${B\to\infty}$, where the resistance (\ref{f1}) simplifies to
\be
\label{res_Maxw}
R_\square=\frac{m}{e^2 \rho_0 \tau}\frac{1}{\ell_R/W  + n_0^2/\rho_0^2}.
\e

The linear behavior follows from the inverse proportionality of the
recombination length to the magnetic field, ${\ell_R\propto1/B}$.  The
saturation occurs when the field becomes so strong, that the ratio
${W/\ell_R}$ becomes comparable with $\rho_0^2/n_0^2$. Clearly, at
charge neutrality ${n_0=0}$ and we recover the unsaturating behavior
of Eq.~(\ref{r2b}).

\subsubsection{General solution}

Having discussed the limiting cases that allow for relatively simple
and physically transparent solutions, we now turn to the most general
case where the two quasiparticle subsystems are characterized by
unrelated masses, momentum relaxation times, equilibrium densities,
and recombination rates. We restrict ourselves to samples with
rectangular geometry and again consider parabolic quasiparticle
spectra (\ref{psp}), arguing that further generalization to arbitrary
spectra will yield no additional physical insight. The main
qualitative conclusion of this section is the same as before: in
classically strong magnetic fields, there exists an intermediate
parameter range, ${\ell_0/\upmu B\ll W\ll\ell_0\upmu B}$ (here $\upmu$
the some averaged mobility of electrons and holes), where the system
exhibits linear magnetoresistance that is non-saturating at charge
neutrality and saturates if the system is doped away from the
neutrality point.

The general solution is most easily obtained upon re-writing the
macroscopic equations (\ref{basic}) in the form similar to
Eqs.~(\ref{basic2}), i.e. in terms of the total quasiparticle density
${\delta \rho=\delta n_e+\delta n_h}$, charge density fluctuations
${\delta n=\delta n_e-\delta n_h}$, quasiparticle flow
${\bb{P}=\bb{j}_e+\bb{j}_h}$, and electric current
${\bb{j}=\bb{j}_h-\bb{j}_e}$. Imposing the hard-wall boundary
conditions (\ref{bc}) on the continuity equation for the electric
current (\ref{div2b}), we find that the lateral component of $\bb{j}$
is equal to zero. All other quantities are functions of the lateral
coordinate $y$. In particular, the macroscopic currents can be written
as
\be
\bb{j}=(j(y),0),
\qquad 
\bb{P}=(P_x(y),P_y(y)).
\e

The first of the equations (\ref{basic}) represents two vector
equations. Re-writing them in terms of the currents $\bb{j}$ and
$\bb{P}$ and writing the resulting equations in components, we obtain
the following four equations
\beml
\label{jPx1}
\be
\label{jnice}
j=\sigma_+E_0+\omega_+P_y,
\e
\be
P_x=\sigma_-E_0+\omega_-P_y,
\e
\beq
\label{nrho2}
(D_+\!+\!\kappa \sigma_+)\frac{\pa  \delta n}{\pa y}
\!+\!D_-\frac{\pa \delta \rho}{\pa y}\!+\!\omega_+P_x\!+\!\omega_-j\!=\!0,
\eq
\beq
(D_-\!+\!\kappa \sigma_-)\frac{\pa \delta n}{\pa y}
\!+\!D_+\frac{\pa \delta \rho}{\pa y}\!+\!\omega_-P_x\!+\!\omega_+j\!=\!0.
\eq
\eml
Here we have used the short-hand notations
\[
\sigma_\pm\!=\!e n_{0,e}\tau_e/m_e\!\pm\! en_{0,h}\tau_h/m_h,
\quad
\omega_{\pm}\!=\!(\omega_e\tau_e\!\pm\!\omega_h\tau_h)/2,
\]
\[
D_\pm \!=\! (D_e\!\pm\! D_h)/2,
\quad
\kappa=e/C,
\]
and took advantage of the gated electrostatics (\ref{electro}) in
order to exclude the lateral component $E_y$.

The second of the equations (\ref{basic}) can be used to obtain the
continuity equation relating the total quasiparticle density
$\delta\rho$ and flow $\bb{P}$, generalizing Eq.~(\ref{div2b}). The
result can be represented in the form
\be
\label{keyE}
\delta \rho=-\frac{1}{\gamma_+}\frac{\pa P_y}{\pa y}
-\frac{\gamma_-}{\gamma_+} \delta n,
\e
where
\[
\gamma_{\pm}=(\Gamma_e\pm\Gamma_h)/4.
\]

Solution of the resulting system of equations (\ref{jPx1}), and
(\ref{keyE}) is straightforward, although tedious. Similarly to the
above solutions of the particular limiting cases, we reduce the
equations (\ref{jPx1}), and (\ref{keyE}) to a single second-order
differential equation, cf. Eqs.~(\ref{basic3}) and (\ref{ddn}).
Here it is convenient to reduce the problem to a second-order
differential equation for $P_y$, 
\be
\label{diffPy}
\frac{\pa^2 P_y}{\pa^2 y}=\frac{4}{\ell_R^2}P_y+\frac{s_0\gamma_-+s_1\gamma_+}{D_0^2} E_0,
\e
where the effective recombination length is given by 
\be
\label{rlgc}
\ell_R =\frac{2D_0}{\sqrt{b_0\gamma_-+b_1\gamma_+}},
\e
and the following notations are introduced
\[
s_0=(\sigma_+\omega_-\!+\!\sigma_-\omega_+)D_+ \!-\! (\sigma_+\omega_+\!+\!\sigma_-\omega_-)D_-,
\]
\[
s_1\!=\!(\sigma_+\omega_+\!+\!\sigma_-\omega_-)(D_+\!+\!\kappa\sigma_+)
\!-\!(\sigma_+\omega_-\!+\!\sigma_-\omega_+)(D_-\!+\!\kappa\sigma_-),
\]
\[
D_0\!=\!\sqrt{D_+(D_+\!+\!\kappa\sigma_+)\!-\!D_-(D_-\!+\!\kappa\sigma_-)},
\]
\[
b_0= 2\omega_+\omega_-D_+ \!-\!(1\!+\!\omega_+^2\!+\!\omega_-^2)D_-,
\]
\[
b_1=(1\!+\!\omega_+^2\!+\!\omega_-^2)(D_+\!+\!\kappa \sigma_+)
\!-\!2\omega_+\omega_-(D_-\!+\!\kappa\sigma_-).
\]
Solving the differential equation (\ref{diffPy}) with the hard-wall
boundary conditions ${P_y(\pm W/2)=0}$ [cf. Eq.~(\ref{bc})], we average
the result over the $y$ coordinate [cf. Eq.~(\ref{Rsquare})],
\be
\overline{P_y}\equiv \frac{1}{W} \int\limits_{-W/2}^{W/2}\!\!\! dy\; P_y(y),
\e
and obtain the solution
\be
\label{zzz}
\overline{P_y}=E_0\frac{s_0\gamma_-+s_1\gamma_+}{b_0\gamma_-+b_1\gamma_+}
\big[F(W/\ell_R)-1\big].
\e
Again, we have used the notation ${F(x)=\tanh(x)/x}$.

\begin{figure}
\centerline{
\includegraphics[width=\columnwidth]{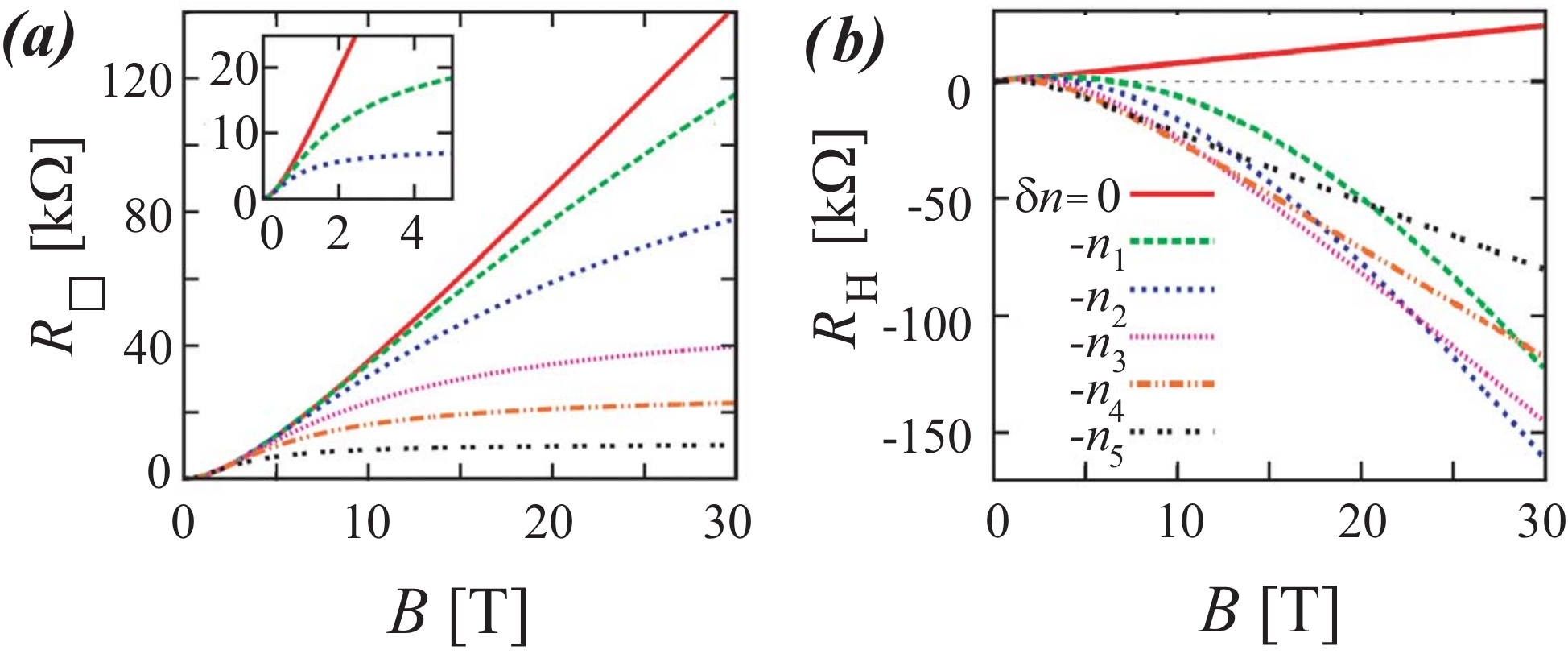}
}
\caption{The magnetic field dependence of the sheet longitudinal
  (left) and Hall resistance (right) given by Eqs.~(\ref{finRs}) and
  (\ref{finRH}). The numerical values were obtained using the typical
  experimental parameters of topological-insulator films: electron and
  hole mobilities ${\upmu_e=20\upmu_h}$,
  ${\upmu_h=1}$~m$^2/$(V$\cdot$s) and velocities ${v_e=10^6}$~m$/$s,
  ${v_h=0.5v_e}$, and the sample parameters are ${W=10\mu}$m,
  ${d=0.5\mu}$m, and ${\epsilon=5}$. The carrier densities were
  calculated using a generic two-band model with the energy gap
  ${\Delta=4}$~meV at room temperature ${T=300}$~K. The recombination
  length in the absence of magnetic field at charge neutrality is
  ${\ell_0=0.37\mu}$m. The solid line corresponds to charge
  neutrality, ${\delta n=0}$, while the other lines correspond to
  negative densities ${\delta n=0.3,0.5,0.9,1.3,2.1\times
    10^{11}}$~cm$^{-2}$. The inset in the left panel shows the
  magnetoresistance $R_{\square}$ for a symmetric model with
  ${\upmu_{\alpha}=20}$~m$^2/$(V$\cdot$s).}
\label{fig:r0rh}
\end{figure}

Finally, we average the relation (\ref{jnice}) in order to find the
average electric current, ${\overline{J}=-e\overline{j}}$, and use the
definition (\ref{Rsquare}) in order to find the inverse sheet
resistance
\be \label{finRs}
R_\square^{-1}\!=\!e\!\lt[\sigma_+\!+\!\omega_+\big[F(W/\ell_R)\!-\!1\big]
\frac{s_0\gamma_-\!+\!s_1\gamma_+}{b_0\gamma_-\!+\!b_1\gamma_+}\rt].
\e
Averaging the $y$-component of the electric field
[cf. Eq.~(\ref{electro})], we find
\[
\overline{E_y}=-\kappa\overline{\frac{\pa \delta  n}{\pa y}}
=\kappa \frac{s_0E_0+b_0\overline{P_y}}{D_0^2}=\eta\,E_0,
\]
with
\[
\eta=\frac{\kappa}{D_0^2}\lt[s_0+b_0\big[F(W/\ell_R)-1\big]
\frac{s_0\gamma_-+s_1\gamma_+}{b_0\gamma_-+b_1\gamma_+}\rt].
\]
As a result, the Hall resistance of the sheet is given by
\be
\label{finRH}
R_\square^{H}=\overline{E_y}/\overline{J}
= \eta\,R_\square.
\e
The results of Eqs.~(\ref{finRs}) and (\ref{finRH}) are shown in
Fig.~\ref{fig:r0rh}, where we plot our results using realistic
parameters for topological-insulator films.

The field dependence of the resistance (\ref{finRs}) comes from the
recombination length (\ref{rlgc}) as well as from the explicit
dependence on the parameters $s_i$ and $b_i$ . However, for
classically strong fields, ${\omega_{\pm}\gg1}$, the latter dependence
cancels out since in this limit the quantities $s_i$ are proportional
to the magnitude of the field, ${s_i=S_iB}$, while $b_i$ are
proportional to the square of the field, ${b_i=B_iB^2}$. The
proportionality coefficients
\[
S_0\!=\!(\sigma_{+}\upmu_{-} \!+\! \sigma_{-} \upmu_{+} ) D_{+}
\!-\!(\sigma_{+} \upmu_{+} \!+\! \sigma_{-} \upmu_{-} ) D_{-},
\]
\[
S_1\!=\!(\sigma_{+}\upmu_{+}\!+\!\sigma_{-}\upmu_{-})(D_{+}\!+\!\kappa\sigma _{ +})
\!-\!(\sigma_{+}\upmu_{-}\!+\!\sigma_{-}\upmu_{+})(D_{-}\!+\!\kappa\sigma _{-}),
\]
\[
B_0\!=\!2 \upmu_{+}\mu_{-} D_{+}\!-\! \left(\upmu_{+}^2 \!+\! \upmu_{-} ^2\right)D_{-},
\]
\[
B_1 \!=\! \left(\upmu_{+}^2 +\upmu_{-} ^2 \right) (D_{+} \!+\! \kappa \sigma _{+})
\!-\! 2 \upmu_{+}\upmu_{-} (D_{-} \!+\! \kappa \sigma _{-}),
\]
are expressed in terms of the effective mobilities
\[
\upmu_{\pm} \!=\! \frac{e}{2c} \left( \frac{\tau_e}{m_e} \!\pm\!
\frac{\tau_h}{m_h} \right).
\]
Then the resistance (\ref{finRs}) becomes similar to all of the above
results (\ref{final0}), (\ref{J}), (\ref{RBolt}), and (\ref{f1}),
insofar the field dependence is confined to the recombination length
$\ell_R$ in the argument of the function ${F(W/\ell_R)}$. Hence, also
the general solution exhibits the three parameter regimes of a
``wide'', ``narrow'', and ``intermediate-sized'' sample. In the most
interesting latter case, the dependence of the resistance
(\ref{finRs}) on the magnetic field may be illustrated by considering
the formal limit ${B\to\infty}$. Then the result can be expressed in the
following simple form,
\be
\label{asimpt_finRs}
R_\square^{-1}=e \left[\sigma_{0} + \frac{M}{B}\right],
\e
where
\begin{subequations}
\label{sigma0_M}
\begin{equation}
\sigma_{0}
=
\sigma_{+}
-
\frac{\upmu_{+} ( S_0 \gamma_{-} + S_1 \gamma_{+}  )}
{ B_0 \gamma_{-} + B_1 \gamma_{+} },
\end{equation}
\begin{equation}
M= \frac{2D_0 \sigma_{+} (S_0 \gamma_{-} + S_1 \gamma_{+} ) }
{ W\, (B_0 \gamma_{-} + B_1 \gamma_{+} )^{3/2}}.
\end{equation}
\end{subequations}
At the neutrality point, where ${n_{0,e}=n_{0,h}=\rho_0/2}$, the
parameters of the solution simplify as
${\sigma_{\pm}=\rho_0\upmu_{\pm}/2}$, and thence
${S_{0(1)}=\rho_0B_{0(1)}/2}$. As a result, ${\sigma_0=0}$, and we
recover the non-saturating LMR.

The results of the previous sections can be obtained from
Eqs.~(\ref{finRs}) and (\ref{asimpt_finRs}) by taking the appropriate
limits. For example, close to charge neutrality, the quantity
$\sigma_0$ is determined by the equilibrium charge density both in the
cases of the electron-hole symmetry and fast Maxwell relaxation
(${\kappa\to\infty}$). In the limit ${n_0\to0}$, we find
${\sigma_0\propto n_0^2}$ and hence
\begin{equation}
R_{\square} ^{-1} =
\sigma_+ \left(\xi + \ell_R/W\right),
\end{equation}
where ${\xi=n_0^2/\rho_0^2}$, similar to Eqs.~(\ref{RBolt}) and
(\ref{res_Maxw}).

\subsection{Hydrodynamic approach}

When the shortest time scale in the problem is due to electron-electron
interaction, 
\be
\label{AA1}
\tau_{\rm ee}^{-1} \gg \tau^{-1}_{\rm imp},~ \tau_{\rm ph}^{-1},
\e
electronic transport may be described using the universal hydrodynamic
approach.

The standard derivation of the hydrodynamic equations relies on the
assumption of local equilibrium, which in a two-component system
could be described by the distribution function
\be
\label{A2}
f_{\alpha}=\frac{1}{\exp\lt\{\lt[\ep_\alpha({\bb{p}})- \bb{p}
    \bb{u}_{\alpha}(\bb{r}) - \mu_{\alpha}(\bb{r})\rt]/T(\bb{r})\rt\}
  +1}.
\e
The electronic fluid is characterized by the local temperature
$T(\bb{r})$, chemical potentials $\mu_\alpha(\bb{r})$, and drift
velocities $\bb{u}_\alpha(\bb{r})$. This distribution function
nullifies the electron-electron and hole-hole collision integrals, but
not the electron-hole collision integral. This means that the standard
approach can only be used if the coupling between the two types of
charge carriers is relatively weak. This is what happens, for example, in
double-layer systems \cite{Narozhny}, where the two types of carriers
belong to physically different layers of the sample.

A complete analytic solution of the kinetic equation of a generic
two-component system with an arbitrary spectrum is not known.  The
problem can be solved in a Fermi liquid \cite{Brooker1972}, but the
resulting theory is rather cumbersome. At the same time, the final
form of the hydrodynamic equations, especially within linear response
\cite{hydrolin}, is universal and is typically believed to be
applicable to most experimentally accessible systems. Here we consider
an electron-hole symmetric [${\ep_{\alpha}(\bb{p})=\ep_{\bb{p}}}$]
system at charge neutrality under a model assumption
\be
\label{hass}
\tau_{\rm eh}^{-1} \ll \tau_{\rm hh}^{-1}=\tau_{\rm ee}^{-1}.
\e
In this case, the equilibration within each subsystem is much faster
than their mutual scattering, so that we can use the distribution
function \eqref{A2} as a starting point. Furthermore, we expect that
even if ${\tau_{\rm eh}\sim\tau_{\rm hh}\sim\tau_{\rm ee}}$ the
effective hydrodynamic description remains valid and describes the
physics of the system at least qualitatively. Remarkably, in graphene
\cite{hydro1,hydrolin,Kashuba2008,Fritz2008,fos,Schuett2011,ryz,svi}
one can rigorously show that the hydrodynamic approach yields a good
quantitative description of electronic transport despite the fact that
the ineqiuality (\ref{hass}) is violated.

For simplicity, we will
assume the parabolic spectrum and energy-independent impurity
scattering time.  Generalization to a more general situation is
straightforward.

Within linear response, the distribution function \eqref{A2}
may be expanded as
\be
\label{A5}
f^\alpha=f^{(0)}+\delta f^\alpha,
\e
\be
\label{A6}
\delta f^\alpha = -\frac{\pa f^{(0)}}{\pa \ep}  \lt(\delta \mu_\alpha +  \ep_{\bb{p}}\,
\frac{\delta T}{T}  + \bb{p}\, \bb{u}_\alpha \rt),
\e
where $\delta\mu_\alpha$, $\delta T$, and $\bb{u}_\alpha$ are
proportional to the electric field $\bb{E}$.

Similarly to the disorder-dominated regime discussed in Sec~\ref{ddr},
we assume here that thermalization between the electronic system and
the lattice is much faster than quasiparticle recombination (even
though both processes are ultimately due to electron-phonon
scattering)
\[
\tau_{\rm ph} \ll \tau_R.
\]
This allows us to neglect local temperature fluctuations
\[
\delta T(\bb{r})=0.
\]
In this case, electrons and hole densities are related to fluctuations
of the chemical potential, $\delta \mu_\alpha$, by means of
Eq.~\eqref{nabla-n}, while the currents are proportional to
hydrodynamic velocities
\be
\bb{j}_\alpha= m\langle v^2\rangle \bb{u}_\alpha/2=\la \ep-\Delta/2 \ra \bb{u}_\alpha.
\e
We remind the reader, that the averaging over all single-particle
states within a given band as defined in Eq.~(\ref{average}) is not
dimensionless. The resulting averaged quantity has dimensions of the
original quantity divided by an extra dimension of energy, such that
the expression ${\langle\ep\rangle}$ is dimensionless.

Usually, the hydrodynamic equations are derived by
multiplying the kinetic equation \eqref{A4} by symmetry-related
factors and integrating over all single-particle states. In
particular, integrating the kinetic equation itself (i.e. with the
factor of unity) yields the continuity equations manifesting the
particle number conservation. In two-component systems, the continuity
equations (\ref{cont-eh}) contain extra factors reflecting
quasiparticle recombination. Integration of the kinetic equation
multiplied by the quasiparticle velocities leads to the macroscopic
equations for the quasiparticle current flows
\beml
\label{currents-eh1}
\beq
\label{jh}
\!\!\!\!\!\!D\bb{\nabla}\delta n_h \!-\!e \bb{E} \rho_0\tau/(2m)
\!-\! \bb{j}_h\times \bb{\omega_c} \tau \!-\!\bb{F}_{eh} \!=\!-\!{\bb{j}_h},
\eq
\beq
\label{je}
\!\!\!\!\!\!D \bb{\nabla} \delta n_e \!+\! e \bb{E} \rho_0\tau/(2m)
\!+\! \bb{j}_e\times \bb{\omega_c} \tau \!+\!\bb{F}_{eh} \!=\!-\!{\bb{j}_e},
\eq
\eml
which differ from Eq.~\eqref{currents-eh} only by the presence of the
friction force 
\be
\bb{F}_{eh}=\chi (\bb{j}_e-\bb{j}_h)/2,
\e
where ${\chi\simeq\tau/\tau_{\rm eh}}$. Under our assumption
(\ref{hass}), the parameter $\chi$ is necessarily small, even though
in the hydrodynamic regime ${\tau/\tau_{\rm ee}\gg1}$ and
${\tau/\tau_{\rm hh}\gg1}$.

At charge neutrality, the currents and densities for the two
quasiparticle branches are not independent for the electron-hole
symmetry dictates the following relations: ${\delta n_h=\delta
  n_e=\delta\rho/2}$, ${j^x_e=-j^x_h=j/2}$, and
${j^y_e=j^y_h=P/2}$. Hence, the continuity equations (\ref{cont-eh})
and macroscopic equations (\ref{currents-eh1}) may be re-written in
the form
\beml
\label{system}
\beq
&&
\label{sys1}
e E \rho_0\tau/m-(1+\chi)j+\omega_c\tau P=0,
\\
&&
\nonumber\\
&&
\label{sys2}
D \pa\delta\rho/\pa y+P+\omega_c\tau j=0,
\\
&&
\nonumber\\
&&
\pa P/\pa y= - \delta\rho/\tau_R,
\eq
\eml
Solving the above equations with the hard-wall boundary conditions
(\ref{bc}), which imply ${P(\pm W/2)=0}$, we find
\beml
\beq
\label{n-beta}
&&
\delta \rho =-\frac{eE_0\ell_R\rho_0\tau}{4m} 
\frac{\omega_c\tau}{D(1+\chi)} \frac{\sinh(2y/\ell_R)}{\cosh(W/\ell_R)}, 
\\
&&
\nonumber\\
&&
\label{jav}
R_\square==\frac{m(1+\chi)}{e^2\rho_0\tau}
\frac{1+\chi+\omega_c^2\tau^2}{1+ \chi+ \omega_c^2\tau^2 F(W/\ell_R)},
\eq
\eml
where the effective recombination length is given by
\be
\ell_R=2\sqrt{\frac{(1+\chi) D\tau_R}{1+\chi +\omega_c^2\tau^2}}.
\e

The dependence of the resistance (\ref{jav}) on the magnetic field is
once again controlled by the function ${F(W/\ell_R)}$. Similarly to the
above discussion of the general disorder-dominated sample, we
illustrate the behavior of the resistance in the
``intermediate-sized'' sample in classically strong magnetic fields by
formally taking the limit ${B\to\infty}$, which here means
${\omega_c\tau\gg\sqrt{1+\chi}}$ and ${W\gg\ell_R}$. In this case we
again find the linear behavior
\be
R_\square=\frac{\sqrt{1+\chi}}{2e\rho_0\sqrt{D\tau_R}}\,B.
\e

The results of this section are qualitatively similar to those
previously obtained in the disorder-dominated regime. In particular,
the resistance (\ref{jav}) differs from Eq.~(\ref{final0}) by the
presence of the parameter $\chi$ describing the mutual friction
between the two carrier subsystems. The friction slightly modifies the
equation for the electric current (\ref{sys1}) as compared to
Eq.~(\ref{n2a}), while the continuity equations and the equation for
the total quasiparticle flow (\ref{sys2}) remain the same as
Eqs.~(\ref{div2b}) and (\ref{n3a}), respectively. This gives us
confidence, that the equations (\ref{system}) provide us with a
general description of electronic transport in two-component systems
close to charge neutrality. Even though the derivation carried out in
this section relied on the simplified model assumption (\ref{hass}),
the resulting equations (\ref{system}) will remain valid for any value
of the electron-hole scattering rate ${1/\tau_{\rm eh}}$.

\section{Transport theory of 3D two-component systems}
\label{sec:3D}

Now we turn to the study of magnetoresistance in 3D two-component
systems. Our goal is to demonstrate, that within the ``classical''
range of magnetic fields, the physics of the phenomenon remains the
same as in the 2D case discussed above. However, practical
calculations are in general difficult. The two main reasons for the
difficulties are (i) the need to solve the 3D Poisson equation to
account for the sample electrostatics, and (ii) a large number of
parameter regimes characterized by competing length scales related to
the sample geometry and microscopic details of the charge carriers, as
well as possible spatial orientations of the applied magnetic field.

\begin{figure}
\centerline{
\includegraphics[width=\columnwidth]{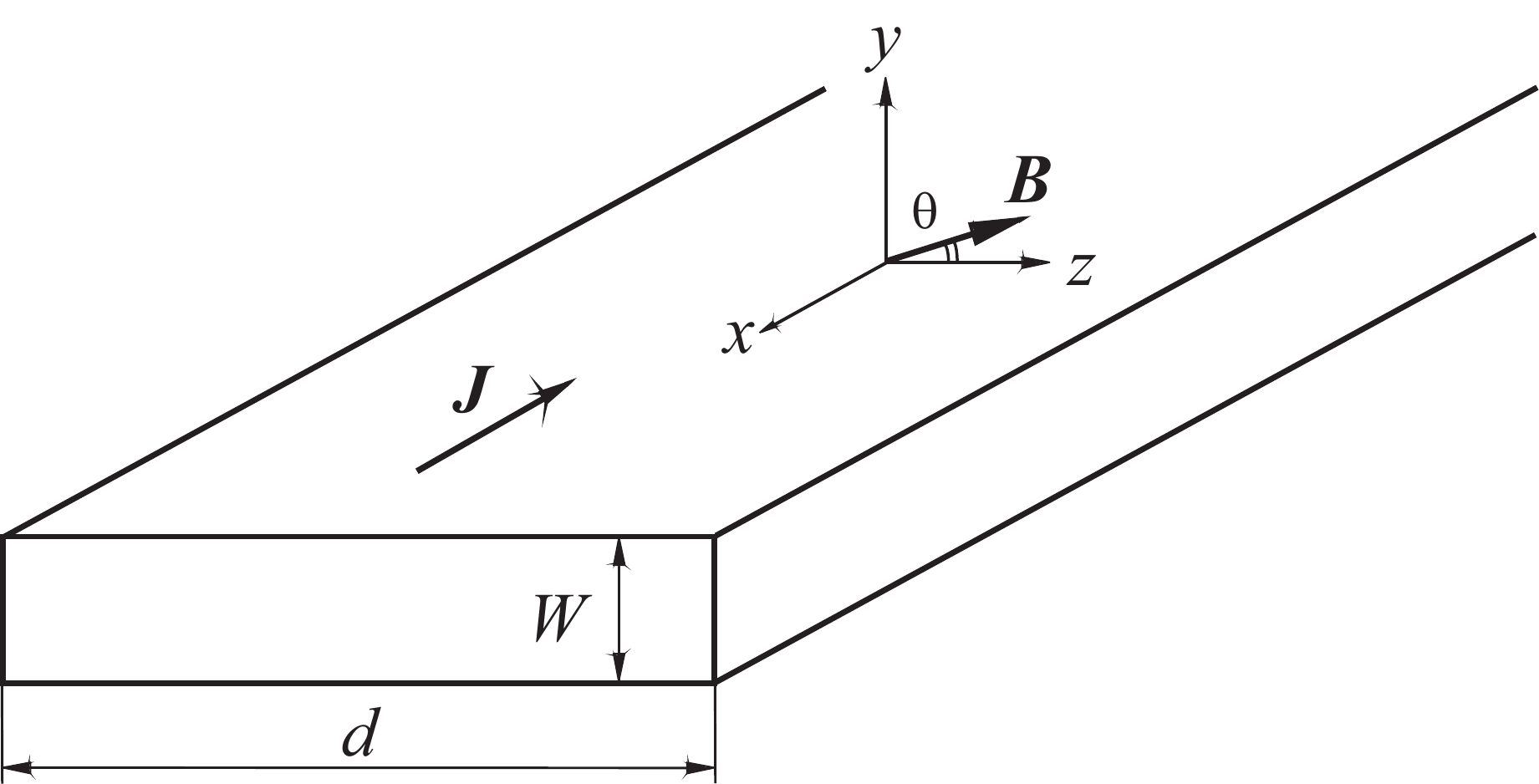}
} \caption{ 3D sample with the shape of a slab in an oblique
magnetic field. The magnetic field vector lies in the $xy$ plane.}
\label{fig:geom}
\end{figure}

In this paper we try to avoid the technical complications as much as
possible by considering a particular ``rectangular'' sample geometry,
see Fig.~\ref{fig:geom}. We consider a sample in the form of a
``slab'', which is ``infinitely'' long (i.e., much longer than any
characteristic length scale in the problem) in one direction, that we
refer to as $x$-direction, while the lateral cross-section of the
sample has a form of a thin rectangle, with one side being much longer
than another, ${d\gg{W}}$ (but still much shorter than the sample size
in the $x$-direction). This particular shape of the sample allows us to
assume that any transport-related quantity is a function of only one
coordinate, $y$.

We assume that the external electric field is applied along the
$x$-direction, ${\bb{E}=E_0\bb{e}_x}$. Consequently, the electric
current is also flowing in the $x$-direction (assuming the hard-wall
boundary conditions in $y$ and $z$ directions)
\begin{equation}
\label{j_3D}
\bb{j} = j(y) \bb{e}_x.
\end{equation}
The applied magnetic field lies in the $yz$-plane,
\begin{equation}
\label{h3d}
\bb{B} = B 
\begin{pmatrix}
0, &  \sin\theta, & \cos\theta
\end{pmatrix}.
\end{equation}
In what follows, we will first consider a particularly simple case,
where the magnetic field is directed along the $z$-axis (${\theta=0}$)
and then discuss the problem with arbitrary $\theta$, focusing on the
neutrality point.

\subsection{Magnetic field orthogonal to the thin, long face of the cuboid sample}

Let us first consider the technically simpler situation where the
magnetic field is applied along the $z$ direction. In this case, the
classical Hall voltage is generated across the $y$ direction. The
relation between the electric field and quasiparticle flows is given
by the standard Ohm's law [cf. Eq.~(\ref{j1}) in the 2D problem]
\begin{subequations}
\label{flows_3D}
\begin{eqnarray}
&&
ej_h^y = \sigma_{h}^{xy} E_0 + \sigma_{h}^{xx} E_y - e D_{h}^{xx} \frac{d\delta n_h}{dy},
\\
&&
\nonumber\\
&&
ej_e^y = \sigma_{e}^{xy} E_0 - \sigma_{e}^{xx} E_y - e D_{e}^{xx} \frac{d\delta n_e}{dy},
\\
&&
\nonumber\\
&&
ej_h^x = \sigma_{h}^{xx} E_0 - \sigma_{h}^{xy} E_y + e D_{h}^{xy} \frac{d\delta n_h}{dy},
\label{jhx3d}
\\
&&
\nonumber\\
&&
ej_e^x = - \sigma_{e}^{xx} E_0 - \sigma_{e}^{xy} E_y - e D_{e}^{xy} \frac{d\delta n_e}{dy}.
\label{jex3d}
\end{eqnarray}
\end{subequations}
In this section, we do not derive the elements of the conductivity
tensor $\sigma_\alpha^{kl}$ and the diffusion constants
$D_\alpha^{kl}$ from the microscopic theory, but rather treat them as
macroscopic (phenomenological) parameters of the system. Furthermore,
we will assume the usual dependence of $\sigma_\alpha^{kl}$ and
$D_\alpha^{kl}$ on the external magnetic field, see Eqs.~(\ref{D-ave})
(\ref{D-DB}), and (\ref{sdb}).

The quasiparticle flows, $j_{e(h)}^y$, obey the continuity equations 
\begin{subequations}
\label{rec_3D}
\begin{eqnarray}
&&
\frac{d j_h^y}{dy}= - (\Gamma_h\delta n_h + \Gamma_e\delta n_e),
\\
&&
\nonumber\\
&&
\frac{d j_e^y}{dy}= - (\Gamma_h\delta n_h + \Gamma_e\delta n_e),
\end{eqnarray}
\end{subequations}
where $\Gamma_{e(h)}$ are the corresponding recombination rates.

Finally, the 3D Poisson equation yields the relationship between the
electric field in the Hall direction and quasiparticle densities,
\begin{equation}
\label{Poisson_3D}
\frac{d E_y}{dy} = 4 \pi e ( \delta n_h - \delta n_e).
\end{equation}

Combining the above equations (\ref{flows_3D}), (\ref{rec_3D}) and
(\ref{Poisson_3D}), we derive a closed system of second-order
differential equations for the quasiparticle density fluctuations
[cf. Eqs.~(\ref{ddn}) and (\ref{drho})]
\begin{subequations}
\label{K}
\begin{equation}
\frac{d^2}{dy^2}
\begin{pmatrix}
\delta n_h \cr
\delta n_e
\end{pmatrix}
=\widehat{K}^2
\begin{pmatrix}
\delta n_h \cr
\delta n_e
\end{pmatrix},
\end{equation}
where the ${2\times2}$ matrix $\widehat{K}$ is given by
\begin{equation}
\widehat{K}^2 =
\begin{pmatrix}
\displaystyle\frac{\Gamma_h+4\pi\sigma_h^{xx}}{D_h^{xx}} & 
\displaystyle\frac{\Gamma_e-4\pi\sigma_h^{xx}}{D_h^{xx}} \cr\cr
\displaystyle\frac{\Gamma_h-4\pi\sigma_e^{xx}}{D_e^{xx}} & 
\displaystyle\frac{\Gamma_e+4\pi\sigma_e^{xx}}{D_e^{xx}}
\end{pmatrix}.
\end{equation}
\end{subequations}

The above differential equations are subject to the hard-wall boundary
conditions [cf. Eq.~(\ref{bc})]
\begin{subequations}
\label{bc3D}
\begin{equation}
\label{q=0_3D} 
j_\alpha^y(y=\pm W/2)=0.
\end{equation}
The boundary conditions (\ref{q=0_3D}) have to be supplemented by the
vanishing boundary conditions \cite{dau8} for the transversal electric
field $E_y$ (see also Appendix~\ref{ebc})
\begin{equation}
\label{Ey=0_D} 
E_y(y=\pm W/2)=0.
\end{equation}
\end{subequations}

The differential equations (\ref{flows_3D}) - (\ref{K}) with the
boundary conditions (\ref{bc3D}) allow for the formal solution
\begin{subequations}
\label{res3d}
\begin{equation}
\label{res3d-n}
\begin{pmatrix}
\delta n_h \cr
\delta n_e
\end{pmatrix}
=
\frac{E_0}{e} \sinh\widehat{K}y
\left[\widehat{K}\cosh\frac{\widehat{K}W}{2}\right]^{-1}\!\!\!
\begin{pmatrix}
\sigma^{xy}_h/D^{xx}_h \cr
\sigma^{xy}_e/D^{xx}_e
\end{pmatrix},
\end{equation}
\begin{widetext}
\begin{eqnarray}
\label{res3d-E}
E_y = 4\pi E_0
\begin{pmatrix}
1 & -1
\end{pmatrix}
\widehat{K}^{-1}
\left[
\cosh\widehat{K}y \left[\cosh\frac{\widehat{K}W}{2}\right]^{-1}\!\!\!-1
\right]
\widehat{K}^{-1}
\begin{pmatrix}
\sigma^{xy}_h/D^{xx}_h \cr
\sigma^{xy}_e/D^{xx}_e
\end{pmatrix},
\end{eqnarray}
\begin{eqnarray}
\label{res3d-jy}
\begin{pmatrix}
j^y_h \cr
j^y_e
\end{pmatrix}
= -\frac{E_0}{e}
\begin{pmatrix}
\Gamma_h & \Gamma_e \cr
\Gamma_h & \Gamma_e
\end{pmatrix}
\widehat{K}^{-1}
\left[
\cosh\widehat{K}y \left[\cosh\frac{\widehat{K}W}{2}\right]^{-1}\!\!\!-1
\right]
\widehat{K}^{-1}
\begin{pmatrix}
\sigma^{xy}_h/D^{xx}_h \cr
\sigma^{xy}_e/D^{xx}_e
\end{pmatrix},
\end{eqnarray}
while the longitudinal currents $j_\alpha^x$ can be found from
Eqs.~(\ref{jhx3d}) and (\ref{jex3d}). The averaged electric current
(\ref{Rsquare}) is
\begin{eqnarray}
\label{javfm}
&&
\overline{J} = E_0
\left[\sigma^{xx}_h\!+\!\sigma^{xx}_e \!+\! \frac{(\sigma^{xy}_h\!-\!\sigma^{xy}_e)^2}
{\sigma^{xx}_h\!+\!\sigma^{xx}_e}\right]
\\
&&
\nonumber\\
&&
\qquad\qquad\qquad\qquad
+E_0
\begin{pmatrix}
D^{xy}_h-4\pi(\sigma^{xy}_h\!-\!\sigma^{xy}_e) &
D^{xy}_e+4\pi(\sigma^{xy}_h\!-\!\sigma^{xy}_e)
\end{pmatrix}
\widehat{F}\left(\frac{\widehat{K}W}{2}\right)
\begin{pmatrix}
\sigma^{xy}_h/D^{xx}_h \cr
\sigma^{xy}_e/D^{xx}_e
\end{pmatrix},
\nonumber
\end{eqnarray}
\end{widetext}
\end{subequations}
where 
\[
\widehat{F}\left(\frac{\widehat{K}W}{2}\right)
=
\frac{2}{W}\widehat{K}^{-1}\sinh\frac{\widehat{K}W}{2}
\left[\cosh\frac{\widehat{K}W}{2}\right]^{-1}.
\]
In an infinitely wide sample (${W\rightarrow\infty}$), the function
$\widehat{F}$ vanishes, leaving the classical result, see the first
line of Eq.~(\ref{javfm}). This comprises the Drude conductivity in
the absence of the magnetic field and the classical, quadratic
magnetoconductivity.

The solutions (\ref{res3d}) are somewhat tedious. Similarly to
Eqs.~(\ref{ddn}) and (\ref{drho}), the matrix $\widehat{K}$ defines
two characteristic length scales, given by its eigenvalues,
$\kappa_{1(2)}$. To make the discussion physically transparent, we
focus on the two limiting cases.

\subsubsection{Fast Maxwell relaxation}

In the limit of fast Maxwell relaxation, determined by the
inequality (the so-called ``good metal'' condition)
\begin{subequations}
\label{egfm}
\begin{equation}
\label{fm}
4\pi\sigma_\alpha^{xx} \gg \Gamma_{\alpha},
\end{equation}
one of the eigenvalues determines the effective recombination length
[cf. Eq.~(\ref{kappa1})]
\begin{equation}
\label{kappa_1_3D}
\kappa_1^2=\frac{4}{\ell_R^2}=
\frac{(\Gamma_h+\Gamma_e)(\sigma_h^{xx}+\sigma_e^{xx})}
{\sigma_h^{xx} D_e^{xx} + \sigma_e^{xx} D_h^{xx}},
\end{equation}
while the other is related to the Thomas-Fermi screening length
\begin{equation}
\label{kappa_2_3D} 
\kappa_2^2 = 4\varkappa^2=4\pi \left(
\frac{\sigma_{h}^{xx}}{D_{h}^{xx}}+\frac{\sigma_{e}^{xx}}{D_{e}^{xx}}
\right) = 4 \pi e^2
\left(
\frac{\partial n_h}{\partial\mu}+\frac{\partial n_e}{\partial\mu}
\right).
\end{equation}
Here ${\partial
  n_\alpha/\partial\mu=\sigma_\alpha^{xx}/(e^2D_\alpha^{xx})}$ is the
thermodynamic density of states. In a typical situation, where the
conductivities and diffusion coefficients for electrons and holes are
of the same order of magnitude, the condition for the fast Maxwell
relaxation (\ref{fm}) can be re-written in one of the two equivalent
forms
\begin{equation}
\label{fm2}
\kappa_1\ll\kappa_2,
\qquad
\varkappa\ell_R\gg 1.
\end{equation}
In classically strong magnetic fields, ${\omega_\alpha\tau_\alpha\gg{1}}$,
the recombination length is inverse proportional to the field,
\begin{equation}
\label{lbfm}
\ell_R\sim1/B,
\end{equation}
while the Thomas-Fermi screening length is approximately
field-independent.
\end{subequations}

In the limit (\ref{fm2}), the results (\ref{res3d})
simplify. Combining the densities (\ref{res3d-n}) into the charge and
quasiparticle densities, we find near one of the boundaries
(${y\approx{W}/2}$)
\begin{subequations}
\label{final-d}
\begin{eqnarray}
\label{final-n}
&&
e\delta n \!=\! \frac{E_0}{2\varkappa} 
\left[
e^{\!-\!\varkappa(W\!-\!2y)}
\left(\!\frac{\sigma_h^{xy}}{D_h^{xx}}\!-\!\frac{\sigma_e^{xy}}{D_e^{xx}}\!\right)
\right.
\\
&&
\nonumber\\
&&
\qquad
\left.
-
\frac{e^{\!-\!(W\!-\!2y)/\ell_R}}{\varkappa\ell_R}
\frac{\sigma_e^{xy}\sigma_h^{xx}\!+\!\sigma_h^{xy}\sigma_e^{xx}}{\sigma_e^{xx}\!+\!\sigma_h^{xx}}
\!\left(\!\frac{1}{D_h^{xx}}\!-\!\frac{1}{D_e^{xx}}\!\right)\!
\right],
\nonumber
\end{eqnarray}
\begin{eqnarray}
\label{final-rho}
&&
\delta\rho\!=\!\frac{E_0}{e}\!
\left[
\frac{e^{\!-\!\varkappa(W\!-\!2y)}}{2\varkappa}
\frac{D_e^{xx}\sigma_h^{xx}\!-\!D_h^{xx}\sigma_e^{xx}}{D_e^{xx}\sigma_h^{xx}\!+\!D_h^{xx}\sigma_e^{xx}}
\left(\frac{\sigma_h^{xy}}{D_h^{xx}}\!-\!\frac{\sigma_e^{xy}}{D_e^{xx}}\right)
\right.
\nonumber\\
&&
\nonumber\\
&&
\qquad\qquad
\left.
+
\ell_R e^{(W\!-\!2y)/\ell_R}
\frac{\sigma_e^{xx}\sigma_h^{xy}\!+\!\sigma_h^{xx}\sigma_e^{xy}}
{D_e^{xx}\sigma_h^{xx}\!+\!D_h^{xx}\sigma_e^{xx}}
\right].
\end{eqnarray}
\end{subequations}
The results (\ref{final-d}) demonstrate the existence of {\it two}
boundary layers forming in the two-component system: (i) the narrow
(in the present limit of fast Maxwell relaxation) {\it screening
  layer}, see also Appendix, and (ii) the wide {\it
  recombination layer}. The latter is similar to the boundary layer
found above in the 2D systems.

The spatial profile of the lateral electric field near the boundary is
similar to Eq.~(\ref{final-n})
\begin{eqnarray}
\label{final-e}
&&
\!\!\!\!E_y\!=\!E_0 \left[ \frac{\sigma_e^{xy}\!-\!\sigma_h^{xy}}{\sigma_e^{xx}\!+\!\sigma_h^{xx}}
\!+\!
\frac{\pi}{\varkappa^2}e^{\!-\!\varkappa(W\!-\!2y)}
\!\left(\!\frac{\sigma_h^{xy}}{D_h^{xx}}\!-\!\frac{\sigma_e^{xy}}{D_e^{xx}}\!\right)
\right.
\\
&&
\nonumber\\
&&
\qquad
\left.
-
\frac{\pi}{\varkappa^2}e^{\!-\!(W\!-\!2y)/\ell_R}
\frac{\sigma_e^{xy}\sigma_h^{xx}\!+\!\sigma_h^{xy}\sigma_e^{xx}}{\sigma_e^{xx}\!+\!\sigma_h^{xx}}
\!\left(\!\frac{1}{D_h^{xx}}\!-\!\frac{1}{D_e^{xx}}\!\right)\!
\right].
\nonumber
\end{eqnarray}
The two quantities satisfy the Poisson equation (\ref{Poisson_3D}).

Finally, the second line in the averaged current (\ref{javfm}) yields
the linear contribution to the magnetoconductivity, which we attribute
to the surface regions of the sample [as opposed to the classical bulk
  contribution given by the first line in Eq.~(\ref{javfm})]
\begin{equation}
\label{reslmc}
\overline{\sigma}_s \approx 2 \frac{\ell_R}{W}
\frac{(D_h^{xy}\!+\!D_e^{xy})(\sigma_e^{xx}\sigma_h^{xy}\!+\!\sigma_h^{xx}\sigma_e^{xy})}
{D_h^{xx}\sigma_e^{xx}\!+\!D_e^{xx}\sigma_h^{xx}}.
\end{equation}
Here we have assumed ${\tanh\kappa_{1(2)}W\!\approx\!1}$,
corresponding to the intermediate sample widths as discussed in the 2D
case, and used Eq.~(\ref{fm2}) to neglect the contribution of the
second eigenvalue $\kappa_2$. The latter is quadratic in the magnetic
field, but vanishes exactly in any compensated system, similarly to
the classical magnetoresistance [as well as the lateral electric field
  (\ref{final-e}) and the fluctuation of the charge density
  (\ref{final-n})].

As a result, a 3D compensated system in orthogonal magnetic field
exhibits linear magnetoresistance in the limit of fast Maxwell
relaxation similarly to the 2D case.

\subsubsection{Slow Maxwell relaxation}

In very strong magnetic fields the condition (\ref{fm}) for fast
Maxwell relaxation is violated and the results (\ref{final-d}),
(\ref{final-e}), and (\ref{reslmc}) become invalid. Assuming that the
motion of charge carriers remains classical, one may consider the
opposite limit of slow Maxwell relaxation where
\begin{subequations}
\label{egsm}
\begin{equation}
\label{sm}
4\pi\sigma_\alpha^{xx} \ll \Gamma_{\alpha}.
\end{equation}
In this case, the eigenvalues of the matrix $\widehat{K}$ (which again
are labeled such that ${\kappa_2\gg\kappa_1}$) are given by
\begin{equation}
\label{kappa_1_sm}
\kappa_1^2=4\pi
\frac{(\Gamma_h+\Gamma_e)(\sigma_h^{xx}+\sigma_e^{xx})}
{\Gamma_h D_e^{xx} + \Gamma_e D_h^{xx}},
\end{equation}
\begin{equation}
\label{kappa_2_sm} 
\kappa_2^2 = \Gamma_h/D^{xx}_{h} + \Gamma_e/D^{xx}_{e}.
\end{equation}
There are still two length scales characterizing the system. Assuming
that the model parameters describing electrons and holes are of the
same order of magnitude, we may associate the smaller eigenvalue
$\kappa_1$ with the inverse Thomas-Fermi screening length and the
larger eigenvalue $\kappa_2$ with the inverse recombination length:
\[
\kappa_1\sim\varkappa, \qquad
\kappa_2\sim1/\ell_R, \qquad
\varkappa\ell_R\ll1.
\]
\end{subequations}
Repeating the calculation leading to Eq.~(\ref{reslmc}) one would now
conclude that the dominant contribution to the magnetoconductivity is
given by the much wider screening surface layer. However, this
contribution
\begin{eqnarray*}
&&
\overline{\sigma}_{sc} \!\approx\! \frac{2}{\kappa_1W}
\frac{\Gamma_hD_e^{xy}\!-\!\Gamma_eD_h^{xy}\!+\!4\pi(\Gamma_h\!+\!\Gamma_e)
(\sigma_h^{xy}\!-\!\sigma_e^{xy})}
{\Gamma_h D_e^{xx} \!+\! \Gamma_e D_h^{xx}}
\\
&&
\\
&&
\qquad\qquad\qquad
\times
(\sigma_h^{xy}\!-\!\sigma_e^{xy}),
\end{eqnarray*}
vanishes for a compensated system. In this case, the formally weaker
contribution of the recombination surface layer determines the field
dependence of the conductivity
\begin{equation}
\label{res2lmc}
\overline{\sigma}_s\!\approx\!\frac{2}{\kappa_2W}
\frac{(\Gamma_h D_e^{xx}\sigma_h^{xy} \!+\! \Gamma_e D_h^{xx}\sigma_e^{xy})
(D_h^{xy}D_e^{xx}\!+\!D_e^{xy}D_h^{xx})}
{(\Gamma_h D_e^{xx} \!+\! \Gamma_e D_h^{xx})D_e^{xx} D_h^{xx}},
\end{equation}
which is linear in the magnetic field.

We conclude that compensated 3D systems with the
geometry of Fig.~\ref{fig:geom} exhibit linear magnetoresistance
when subjected to the perpendicular magnetic field.

\subsection{Oblique magnetic field}

Consider now the general situation where the magnetic field is
not collinear with any sample edges, see Fig.~\ref{fig:geom}. In this
section we restrict ourselves to the electron-hole symmetric system at
charge neutrality. In this case, the macroscopic equation describing
transport properties of the system can be simplified similarly to the
2D case.

In the geometry of Fig.~\ref{fig:geom} and under the assumption
${d\gg{W}}$, all physical quantities depend only on the coordinate
$y$. Hence the continuity equation for the total quasiparticle flow,
${\bb{P}=(0, P_y(y), P_z(y))}$, takes the form [cf. Eq.~(\ref{div2b})]
\begin{subequations}
\label{bal_eqs_oblique}
\begin{equation}
\label{ce3}
P_y'=-\delta\rho / \tau_R.
\end{equation}
The equations expressing the relation between the quasiparticle flows
and the electric field [cf. Eqs.~(\ref{j1}), (\ref{currents-eh1}), and
  (\ref{flows_3D})] can be expressed in terms of the total
quasiparticle flow and the electric current (\ref{j_3D}):
\begin{equation}
\label{olp3}
D
\begin{pmatrix}
\delta\rho' \cr
0
\end{pmatrix}
+
\begin{pmatrix}
P_y \cr
P_z
\end{pmatrix}
+
j\tau
\begin{pmatrix}
\omega_z \cr
-\omega_y
\end{pmatrix}
=0,
\end{equation}
\begin{equation}
\label{olj3}
j-e E_0 \rho_0 \tau /m   +\tau ( P_z \omega_y - P_y \omega_z ) = 0,
\end{equation}
\end{subequations}
where ${\omega_{y(z)}=eB_{y(z)}/mc}$.

Solving the equations (\ref{bal_eqs_oblique}) with the hard wall
boundary conditions, ${P_y(y=\pm W/2)=0}$, we find
\[
P_y(y)=
\frac{j_0\omega_z \tau}{1+\tau^2 \omega_c^2}
\left[
\frac{\cosh \lambda y }{\cosh (\lambda W/2)}
-1
\right],
\]
and 
\[
j(y)=
\frac{j_0}{1+\tau^2 \omega_c^2}
\left[
1+
\frac{\omega_z^2 \tau^2}{ 1 + \omega_y^2 \tau^2}
\frac{\cosh \lambda y }{\cosh (\lambda W/2)}
\right].
\]
Here ${j_0=eE_0\rho_0\tau/m}$ is the electric current in the absence
of the magnetic field, $\omega_c^2=\omega_y^2+\omega_z^2$, and
$\lambda$ is the inverse field-dependent ``recombination length'',
\[
\lambda ^2 = \frac{1}{D \tau_R} \left[
1+
\frac{\tau^2 \omega_z^2}{1+\tau^2 \omega_y^2}
\right].
\]

The resulting averaged resistance of the sample is calculated
similarly to the case of the orthogonal geometry and has the form
\begin{equation}
\label{Result_oblique}
R_{\square} =\frac{m}{e^2 \rho_0 \tau}
\frac{1+\tau^2 \omega_c^2}{1+
\frac{\tau^2 \omega_z^2}{1+ \tau^2 \omega_y^2}
F(\lambda W/2)},
\end{equation}
where, as defined above, ${F(x)=\tanh x/x}$.

In narrow samples, ${\lambda W\ll 1}$, recombination is ineffective
since the time it takes the carriers to move from one slab facet to
another is smaller than the typical recombination (as well as
diffusion) time. Nevertheless, in contrast to the above case of the
orthogonal geometry the magnetoresistance is nonzero,
\[
R_{\square} = \rho _0(1+\tau^2 \omega_y^2),
\]
and is determined by the $y$-component of the magnetic field. The 
physical reason fro this result is the effect of the magnetic field 
on carrier motion in the $z$-direction.

In wide samples,
\[
\lambda W \gg \tau^2 \omega_z^2/(1+ \tau^2 \omega_y^2),
\]
we recover the classical bulk magnetoresistance, which is quadratic
in the applied magnetic field:
\[
R_{\square}=\varrho _0[1+\tau^2\omega_c^2].
\]
This result can also be obtained within the Drude theory of
two-component systems if the electric current is allowed to flow only
in the $x$-direction.

\begin{figure}
\centerline{\includegraphics[width=\columnwidth]{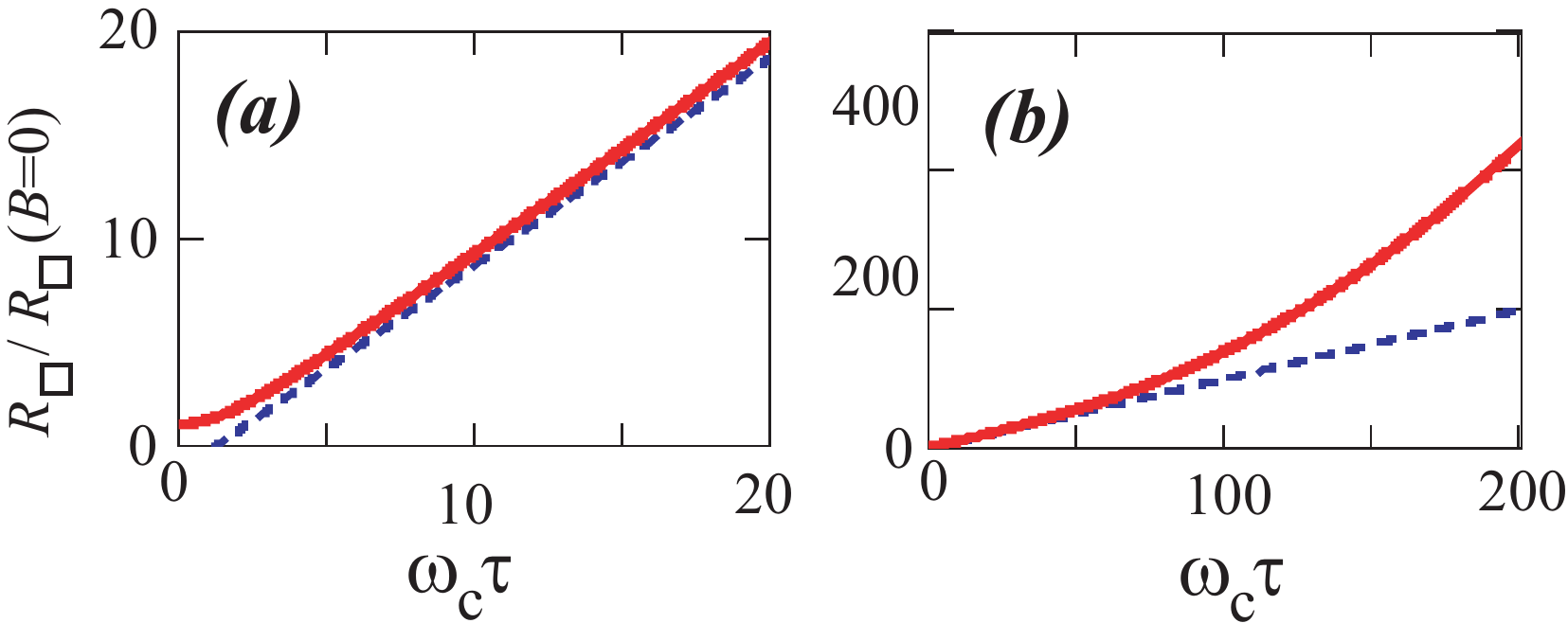}} 
\caption{(Color online) 3D two-component system at charge neutrality
  in an oblique magnetic field. The recombination length in zero
  magnetic field is taken to be equal to the sheet width:
  $W/\ell_0=1$. The solid red curves represent the calculated values
  of the magnetoresistance. The field is directed at the angle
  ${\theta=0.5^o}$. The two panels show the same result in the two
  different ranges of the parameter ${\omega_c\tau}$: the panel (a)
  shows the onset of the intermediate, nearly linear behavior, while
  the panel (b) shows the recovery of the quadratic magnetoresistance
  in strong fields. The dashed blue lines are guides to the eye.}
\label{fig:picture}
\end{figure}

Finally, one may consider the intermediate situation:
\begin{equation}
\label{cr_lin_obl}
1 \ll \lambda W \ll \tau^2 \omega_z^2/(1+ \tau^2 \omega_y^2).
\end{equation}
Such an interval may only exist when the direction of the magnetic
field is almost orthogonal to the sample face: ${\omega_z\gg\omega_y}$
or, equivalently, ${\theta\ll1}$. Then the sample resistance takes the
form:
\begin{equation}
\label{R__edge_oblique}
R_{\square}= \frac{\rho_0 W}{2\sqrt{D \tau_R}}
\frac{[1+\tau^2 \omega_y^2]^{1/2}
[1+\tau^2 \omega_c^2]^{3/2}}
{\tau^2 \omega_z^2}.
\end{equation}
Using ${\omega_z\tau\gg1}$ and ${\omega_y\ll\omega_z}$,
the above expression can be simplified to
\begin{equation}
\label{R__edge_oblique_det}
R_{\square}=\rho_0 (W/\ell_0)
\tau \omega_z \sqrt{1+\tau^2 \omega_y^2}.
\end{equation}
The resistance (\ref{R__edge_oblique_det}) exhibits an approximately
linear field dependence if ${\tau\omega_y\ll1}$, such that the square 
root may be approximate by unity. In a general situation the magnetoresistance
is quadratic, see Fig.~\ref{fig:picture}.

\section{Conclusions}
\label{sec:concl}

In this paper, we have studied the recombination mechanism of
magnetoresistance in finite-size, two-component systems near charge
neutrality \cite{Alekseev15}. Precisely at the neutrality point the
classical Hall effect is compensated. In particular, there is no Hall
voltage. The electric current flowing through the system is
accompanied by a lateral, neutral quasiparticle flow. In any
finite-size system (i.e. in any sample studied in laboratory
experiments) this flow terminates at the boundary leading to
quasiparticle accumulation in the well-defined edge region, see
Fig.~\ref{fig:qualpic}. The width of that region is determined by
inelastic scattering processes and is of the order of the
recombination length. The latter depends on the external magnetic
field and hence the edge region contributes to the overall
magnetoresistance of the sample. The relative strength of this
contribution (as compared to the bulk of the system) depends on the
sample geometry, strength of the recombination processes, and magnetic
field. In strong enough magnetic fields, there exist a wide region of
parameters, where the edge contribution dominates over the bulk
leading to the linear dependence of the sample resistance on the
external field.

Our explicit calculations show that the recombination mechanism of LMR
in compensated two-component systems is generic and independent of the
details of the quasiparticle excitation spectrum. Away from the
neutrality point, the linear field dependence eventually saturates at
the strongest (but still classical) fields. Such strong dependence of
the magnetoresistance on the carrier density distinguishes the
recombination mechanism from the previously proposed extreme quantum
\cite{Abrikosov1969,Abrikosov1998,Abrikosov2000} and classical
\cite{Dykhne,Parish2003,Knap2014} theories.

Magnetoresistance observed in experiments on compensated two-component
systems \cite{Wiedmann2015,Vasileva16} does exhibit the essential
qualitative features of the recombination mechanism. At the same time,
LMR is observed in a wide variety of materials, many of which do not
conform to the assumptions of the present paper. It is therefore very
interesting to extend the theory of recombination-assisted
magnetoresistance in two-component materials to the cases of strongly
disordered systems (including the long-wavelength, smooth disorder),
systems where recombination processes are mostly effective near the
boundaries, and situations where the electron-phonon coupling is not
strong enough to provide a mechanism for fast energy relaxation and
thermalization.

\section*{Acknowledgments}

We thank U. Briskot, Yu. Vasilyev, and B. Yan for fruitful
discussions. The work is supported by the Dutch Science Foundation
NWO/FOM 13PR3118, the EU Network FP7-PEOPLE-2013-IRSES Grant 612624
``InterNoM'', the Russian Federation President Grant MK-8826.2016.2,
the Russian Foundation for Basic Research, and the Russian Ministry of
Education and Science.

\appendix*

\section{Screening boundary layer in the classical Hall effect}
\label{ebc}

Here we discuss the boundary layer in the classical Hall effect in
disordered metals with {\it finite conductivity}. In contrast to the
textbook case of an ideal conductor, here the charges are not confined
to the boundary and the electric field is nonzero inside the metal.

\begin{figure}[t]
\centerline{
\includegraphics[width=0.42\columnwidth]{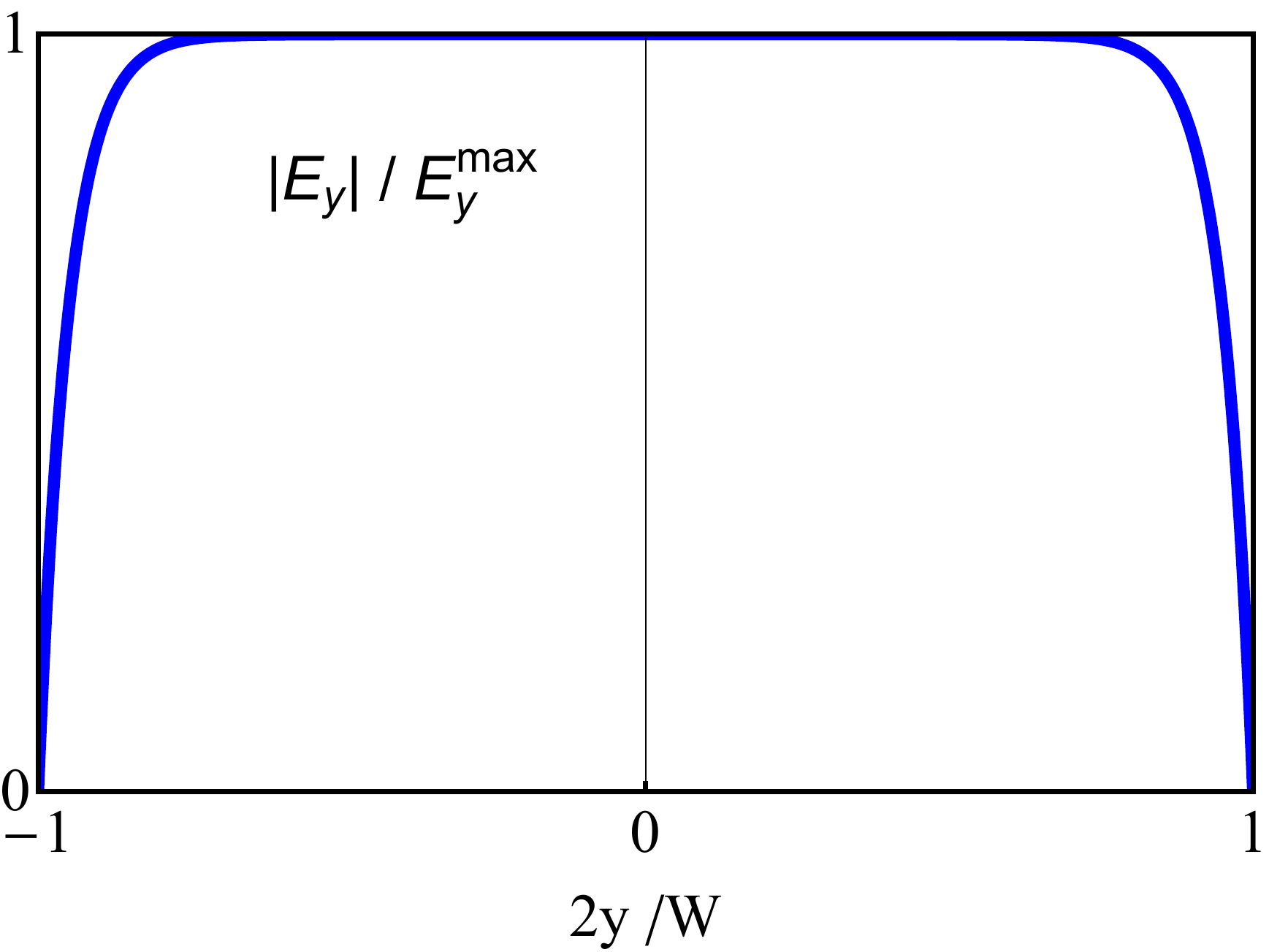}
\qquad
\includegraphics[width=0.42\columnwidth]{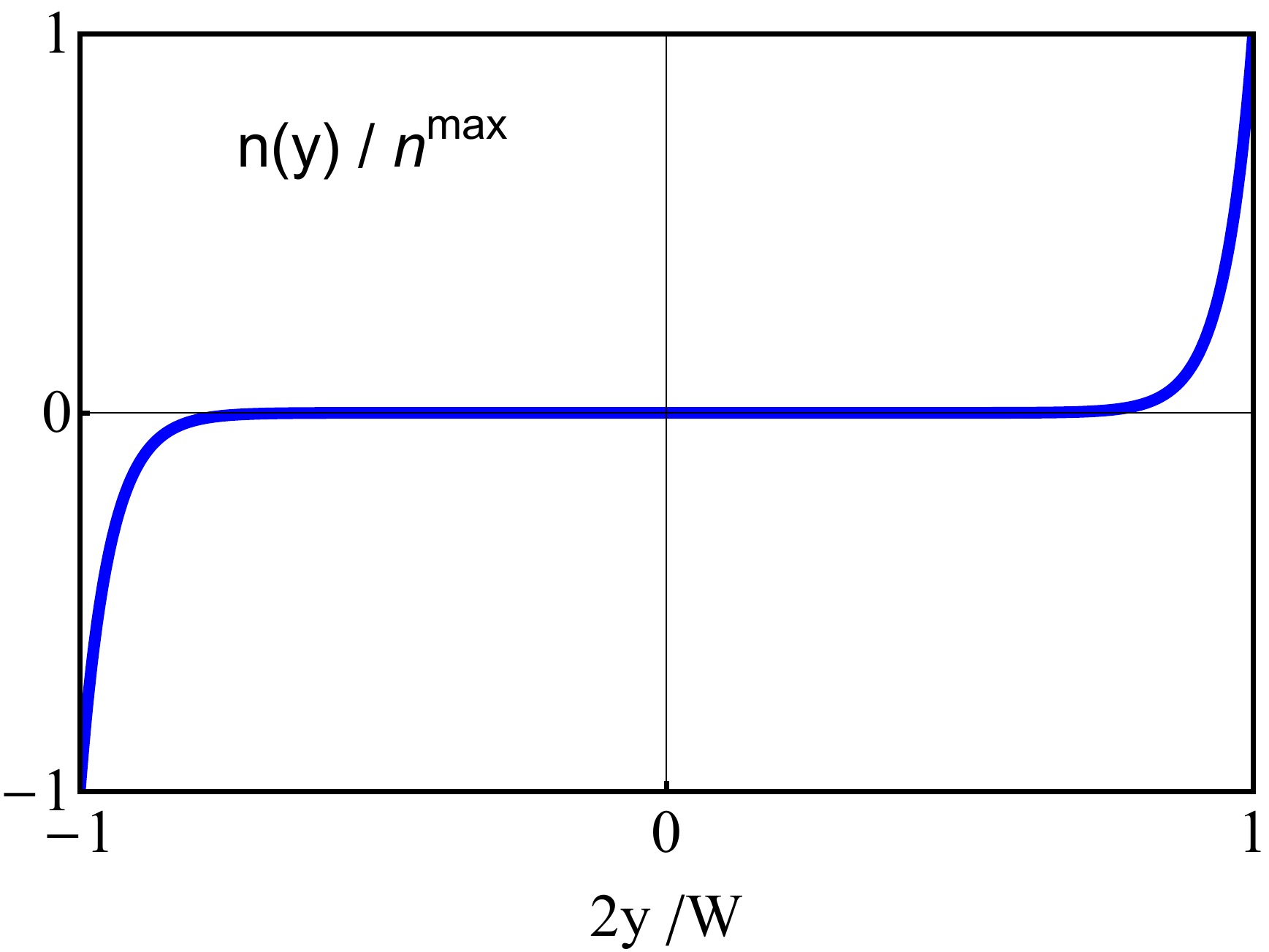}
} 
\caption{Spatial profiles of the lateral electric field (left) and
  charge density (right) in the classical Hall effect in a sample with
  the slab geometry of Fig.~\ref{fig:geom}.}
\label{fig:bc}
\end{figure}

The electric current density, ${\bb{J}}$, inside a metal is related to
the electric field and charge density by means of Ohm's law
[cf. Eq.~(\ref{j1})],
\begin{equation}
\label{ohm}
\bb{J} = \hat\sigma \bb{E} - e \hat D(B) \bb{\nabla} n,
\end{equation}
where $n$ denotes the {\it volume density} of charge carriers, such
that ${en}$ is the charge density. In the presence of the magnetic
field, the diffusion coefficient is represented by the matrix
\[
\hat D(B)  = \frac{D}{1+\omega_c^2\tau^2}
\begin{pmatrix}
1 & \omega_c\tau \cr
\omega_c\tau & 1
\end{pmatrix},
\]
see Eq.~(\ref{D-DB}). In a steady state, we may write the continuity equation as
[cf. Eq.~(\ref{div2b})]
\begin{equation}
\label{con}
\bb{\nabla}\cdot\bb{J}=0.
\end{equation}
Finally, the electric field and charge density are related by
Maxwell's equation,
\begin{equation}
\label{max}
\bb{\nabla}\cdot\bb{E}=4\pi en.
\end{equation}
Taking the gradient of Eq.~(\ref{ohm}) and using Eqs.~(\ref{con}) and
(\ref{max}), one finds
\begin{equation}
\label{grohm}
\bb{\nabla}\cdot\left(\hat\sigma \bb{E}\right) 
= e \bb{\nabla} \cdot\left(\hat D \bb{\nabla} n\right).
\end{equation}
Solution to the coupled differential equations (\ref{max}) and
(\ref{grohm}) depends on the system geometry.

Assuming the simplest geometry of Fig.~\ref{fig:geom}, we may exclude
the electric field from Eqs.~(\ref{max}) and (\ref{grohm}). This way we
find the equation \cite{Shik} for the carrier density ${n(y)}$
[cf. Eqs.~(\ref{ddn}), (\ref{drho}), and (\ref{K})],
\begin{equation}
\label{nd}
n'' = 4\varkappa^2n,
\end{equation}
where $\varkappa$ is the inverse Thomas-Fermi screening length,
${\varkappa=\sqrt{\pi\sigma^{xx}(B=0)/D}}$ (assuming
$\sigma^{yy}=\sigma^{xx}$).

Solving Eq.~(\ref{nd}) with the hard wall boundary conditions
\cite{dau8} [cf. Eqs.~(\ref{bc}) and (\ref{Ey=0_D})]
\begin{equation}
\label{bcje}
J^y(\pm W/2)=0, \qquad
E_y(\pm W/2)=0,
\end{equation}
we find
the carrier density profile
\begin{equation}
\label{nres}
n(y) = \frac{\sigma^{yx}E_0}{2e\varkappa D^{yy}}
\frac{\sinh 2\varkappa y}{\cosh \varkappa W}.
\end{equation}
The lateral component of the electric field is given by
\begin{equation}
\label{eres}
E_y(y)=\frac{\pi\sigma^{yx}E_0}{\varkappa^2D^{yy}}
\left(\frac{\cosh 2\varkappa y}{\cosh \varkappa W} - 1\right).
\end{equation}
These results are illustrated in Fig.~\ref{fig:bc}. Clearly, the
off-diagonal component of the conductivity matrix,
${\sigma^{yx}\propto{B}}$, is nonzero only in the presence of the
external magnetic field. Both the charge density and electric field are
non-uniform close to the sample boundaries. The width of the
corresponding boundary layer is determined by the screening length.
In an ideal conductor, the screening length is equal to zero. In this
limit, the charge density (\ref{nres}) develops a singularity at the
boundary corresponding to the {\it surface} charge density [which in
  turn leads to the jump of the lateral electric field at the boundary
  \cite{dau8} in contrast with the boundary condition (\ref{bcje})].

Substituting the results (\ref{nres}) and (\ref{eres}) into
Eq.~(\ref{ohm}), we recover the usual field-independent resistance,
typical for single-component systems \cite{Pippard,dau8,abrikos}.


\begin{thebibliography}{99}
\bibitem{footnote} 
Note, that LMR may also appear in short samples \cite{Lakeou}, see
also Ref.~\onlinecite{Weber16} as well as the earlier work
\cite{Yoshida}. Such effects are beyond the scope of the present
paper.
\end{thebibliography}
\end{document}